\documentclass[twocolumn,tighten]{aastex63}

\usepackage{amsmath}
\usepackage{amssymb}
\usepackage{ragged2e}
\received{July XX, 2020}
\revised{XXXX XX, 2020}
\accepted{XXX}%\today}
\submitjournal{ApJ}

\shorttitle{Tracing the formation of giant planets with C, O, N and S}
\shortauthors{Turrini et al.}
\begin{document}

\title{Tracing the formation history of giant planets in protoplanetary disks with Carbon, Oxygen, Nitrogen and Sulphur}%\footnote{Released on January, 8th, 2018}}

%% The \author command is the same as before except it now takes an optional
%% arguement which is the 16 digit ORCID. The syntax is:
%% \author[xxxx-xxxx-xxxx-xxxx]{Author Name}
%%
%% This will hyperlink the author name to the author's ORCID page. Note that
%% during compilation, LaTeX will do some limited checking of the format of
%% the ID to make sure it is valid.
%%
\correspondingauthor{Diego Turrini.}
\email{diego.turrini@inaf.it}

\author[0000-0002-1923-7740]{Turrini D.}
\affiliation{INAF - Istituto di Astrofisica e Planetologia Spaziali (INAF-IAPS), Via del Fosso del Cavaliere n. 100, 00133, Rome, Italy}

\author[0000-0003-1560-3958]{Schisano E.}
\affiliation{INAF - Istituto di Astrofisica e Planetologia Spaziali (INAF-IAPS), Via del Fosso del Cavaliere n. 100, 00133, Rome, Italy}

\author[0000-0002-3911-7340]{Fonte S.}
\affiliation{INAF - Istituto di Astrofisica e Planetologia Spaziali (INAF-IAPS), Via del Fosso del Cavaliere n. 100, 00133, Rome, Italy}

\author[0000-0002-9826-7525]{Molinari S.}
\affiliation{INAF - Istituto di Astrofisica e Planetologia Spaziali (INAF-IAPS), Via del Fosso del Cavaliere n. 100, 00133, Rome, Italy}

\author[0000-0002-9793-9780]{Politi R.}
\affiliation{INAF - Istituto di Astrofisica e Planetologia Spaziali (INAF-IAPS), Via del Fosso del Cavaliere n. 100, 00133, Rome, Italy}

\author[0000-0001-6156-0034]{Fedele D.}
\affiliation{INAF - Osservatorio Astrofisico di Arcetri, Largo E. Fermi 5, 50127 Firenze, Italy}

\author[0000-0002-6648-2968]{Pani\'c O.}
\affiliation{School of Physics and Astronomy, E.~C.~Stoner Building, University of Leeds, Leeds LS2 9JT, United Kingdom}

\author[0000-0003-0065-7267]{Kama M.}
\affiliation{Tartu Observatory, University of Tartu, Observatooriumi 1, 61602 T\~{o}ravere, Estonia}
\affiliation{Department of Physics and Astronomy, University College London, London, WC1E 6BT, United Kingdom}

\author[0000-0001-6516-4493]{Changeat Q.}
\affiliation{Department of Physics and Astronomy, University College London, London, WC1E 6BT, United Kingdom}

\author[0000-0001-6058-6654]{Tinetti G.}
\affiliation{Department of Physics and Astronomy, University College London, London, WC1E 6BT, United Kingdom}

\begin{abstract}
{

The composition of giant planets is imprinted by their migration history and the compositional structure of their hosting disks. Studies in recent literature investigate how the abundances of C and O can constrain the formation pathways of giant planets forming within few tens of au from the star. New ALMA observations, however, suggest planet-forming regions possibly extending to hundreds of au. We explore the implications of these wider formation environments through n-body simulations of growing and migrating giant planets embedded in planetesimal disks, coupled with a compositional model of the protoplanetary disk where volatiles are inherited from the molecular cloud and refractories are calibrated against extrasolar and Solar System data. We find that the C/O ratio provides limited insight on the formation pathways of giant planets that undergo large-scale migration. This limitation can be overcome thanks to nitrogen and sulphur. Jointly using the C/N, N/O and C/O ratios breaks any degeneracy in the formation and migration tracks of giant planets. The use of elemental ratios normalized to the respective stellar ratios supplies additional information on the nature of giant planets, thanks to the relative volatility of O, C and N in disks. When the planetary metallicity is dominated by the accretion of solids C/N* $>$ C/O* $>$ N/O* (* denoting this normalized scale), otherwise N/O* $>$ C/O* $>$ C/N*. The S/N ratio provides an additional independent probe into the metallicity of giant planets and their accretion of solids.

} 
\end{abstract}


\section{Introduction}

The formation of giant planets is one of the defining milestones in the life of planetary systems, and plays a fundamental role in shaping their architecture and habitability \citep[e.g.][and references therein]{morbidelli2016}. Understanding the formation pathways of giant planets, however, is proving quite a challenging task, particularly in terms of answering the following questions: where do they form, and how and when do they acquire their final orbits?

The orbital architectures of the known exoplanetary systems, with their large population of giant planets at a fraction of au from the star, provide a strong argument that migration processes play a crucial role in the formation pathways of these planets. A prime mechanism for their migration is generally identified in the interaction and exchange of angular momentum with the protoplanetary disk \citep[e.g.][and references therein]{nelson2018}, that also allows the migrating planet to encounter and capture more significant quantities of solid material than it would otherwise \citep[e.g.][and references therein]{turrini2015,shibata2019,shibata2020}. 

Population studies of the metallicities of giant planets, both around other stars \citep{thorngren2016,wakeford2017} and in our Solar System \citep[e.g.][and references therein]{atreya2018}, have revealed widespread super-stellar enrichments in high-Z elements in their H/He-dominated gaseous envelopes, consistent with the coupling between early disk-driven migration and enhanced accretion of solid material. The highest estimated metallicities, in particular, can currently be explained only by invoking extensive disk-driven migration from the outer regions of protoplanetary disks, coupled with equally extensive accretion of planetesimals \citep{shibata2020}. 

This scenario appears to be supported by the possible large extension of the planet-forming region in protoplanetary disks suggested by recent observations with ALMA \citep[e.g.][and references therein]{alma2015,isella2016,fedele2017,fedele2018,long2018,andrews2018}. These observations revealed the widespread presence of morphological features ascribed to the presence of giant planets from several tens to a few hundreds of au from their central stars. However, those very observations also revealed that giant planets could remain at such distances from their stars for several millions of years, raising the question of whether disk-driven migration is a common occurrence or not.

Furthermore, the architecture of a large fraction of exoplanetary systems shows signatures of a chaotic dynamical evolution \citep{limbach2015,zinzi2017,laskar2017,turrini2020}, suggesting that not all giant planets have their dynamical histories shaped by early disk-driven migration, but they can reach their final orbits at a later time as a result of planet-planet scattering events, chaotic dynamics or a combination of early and late migration. Due to these uncertainties , metallicity alone is not enough to fully disclose a giant planet's formation history, aside from the most extreme cases. %Late migration events have smaller impacts on the envelope metallicity, as fully-formed giant planets are more efficient in scattering away planetesimals rather than accreting them \citep[e.g.][and references therein]{turrini2015,shibata2019}.

%Due to these uncertainties, metallicity alone is not enough to disclose the details of \textcolor{red}{a giant planet's} formation history\textcolor{red}{\textst{ies}}, aside from the most extreme cases. %As an example, two giant planets could show similar metallicity values and share similar final orbits\textcolor{red}{. However,} they could have experienced completely different combinations of early and late migration, with their disk-driven migration and accretion of planetesimals having occurred in completely unrelated regions of their disks.

The information provided by the atmospheres of giant planets on their interior composition allows for breaking this degeneracy \citep[see e.g.][and references therein for recent discussions]{madhusudhan2016,tinetti2018,turrini2018,madhusudhan2019}. In the coming years, the task of characterizing in detail the atmospheric composition of exoplanets will be first undertaken by the \textit{James Webb Space Telescope} \citep[\textit{JWST},][]{cowan2015}. Soon after, the space mission \textit{Ariel} of the European Space Agency will address this task in a systematic way for a sample of several hundreds of exoplanets \citep{tinetti2018}.
%\textcolor{red}{. It will} be later more systematically addressed by the space mission Ariel \citep{tinetti2018} of the European Space Agency.

Since the early work of \citet{oberg2011}, several studies have been devoted to understanding the link between the abundances of the two most abundant high-Z elements, carbon (C) and oxygen (O), and the planetary formation process \citep[see e.g.][and references therein for recent discussions]{madhusudhan2016,mordasini2016,booth2019,cridland2019}. Recently, a few studies have taken a first look at nitrogen (N), as well \citep[][]{oberg2019,bosman2019}. However, the implications of formation regions extending as far from the star as ALMA reveals, as well as those of accreting large amounts of high-Z material across many of the compositional regions inside disks, remain poorly understood.

The goal of this study is, therefore, twofold. On the one hand, we want to investigate the information provided by the C/O ratio when the migration tracks of giant planets extend over tens or hundreds of au and cross many of the diverse  compositional regions in protoplanetary disks. On the other hand, we want to expand our suite of compositional tracers to nitrogen (N) and sulphur (S) to improve our understanding of the link between planetary composition and formation process.

The rest of this work is organized as follows. In Sect. \ref{sect-model} we detail the dynamical and compositional models at the basis of our study. In Sect. \ref{sect-results} we present the outcomes of the investigated scenarios in terms of planetary metallicity and abundance ratios of the four elements under study. In Sect. \ref{sect-discussion} we discuss the additional information that is provided by the combined use of planetary and stellar elemental ratios, as well as the implications of the uncertainty on the volatility of N in protoplanetary disks. Finally, in Sect. \ref{sect-conclusions}
we draw the conclusions of our work and summarise the applications of our results.
 
%A growing body of work has been investigating the compositional features created by the formation environments and by migration on planets and how they can be used to understand their formation histories \citep[e.g.]{oberg2011,johnson2012,thiabaud2014,marboeuf2014a,marboeuf2014b,mordasini2016,cridland2019}.  In particular, the abundance ratio between two of the most cosmically abundant elements, oxygen (O) and carbon (C), has been identified as a potential window into the formation region of giant planets since the early work of \citet{oberg2011},

\section{Numerical and compositional model}\label{sect-model}

Our work is based on two complementary modelling efforts. The first one is a set of detailed n--body simulations, each tracking the dynamical evolution of planetesimals in response to the growth and migration of a forming giant planet, sampling different formation regions in the protoplanetary disk. The second is a compositional model for the solids and the gas in the protoplanetary disk, accounting for the information provided by both solar and extrasolar materials. The n-body simulations were performed with the {\sc Mercury-Ar$\chi$es} software \citep{turrini2019}, briefly described in Sect \ref{sect-nbody_code}. Details on the treatment of the physical processes included in the n-body simulations are provided in Sects. \ref{sect-giant_planets}, \ref{sect-protoplanetary_disk}, and \ref{sect-planetesimal_disk}. The compositional model used to analyze the outcomes of the n-body simulations is described in Sect. \ref{sect-compositional_model}.

\subsection{{The n-body code \sc Mercury-Ar$\chi$es}}\label{sect-nbody_code}
To simulate the formation and migration of the giant planet and its interactions with the planetesimal disk, we used {\sc Mercury-Ar$\chi$es} \citep{turrini2019}, a high-performance implementation of the hybrid symplectic algorithm of the {\sc Mercury} 6 software from \citet{chambers1999}. Symplectic algorithms \citep{wisdom1991,kinoshita1991} are a family of leapfrog integration schemes that are well-suited for simulating the evolution of planetary systems over long timescales, as they show no long-term accumulation of the energy error while being about an order of magnitude faster than conventional n-body algorithms.\\
\indent The energy conservation property of symplectic algorithms is guaranteed as long as the hierarchical structure of the dynamical system is not violated: for planetary systems this means that the dynamical evolution of the bodies should be dominated by the gravitational field of the central star. This condition is violated during close encounters between planetary bodies, in principle making symplectic algorithms unsuitable for studies of planetary formation.\\
\indent Hybrid symplectic algorithms allow to circumvent this problem by switching to high-precision non-symplectic integration schemes for the duration of the close encounter in reproducing the dynamical evolution of the involved bodies \citep{chambers1999}. To guarantee high numerical stability and accuracy, {\sc Mercury}'s hybrid symplectic algorithm uses the Bulirsh-Stoer integration scheme to compute the keplerian motion of the bodies undergoing close encounters \citep{chambers1999}. {\sc Mercury-Ar$\chi$es} implements a high-performance version of the hybrid symplectic scheme of {\sc Mercury}, parallelized and vectorized with OpenMP to take advantage of the SIMD (single instruction, multiple data) capabilities of modern multi-core architectures.\\
\indent {\sc Mercury-Ar$\chi$es} further expands the hybrid symplectic algorithm of {\sc Mercury} with the inclusion of the following effects on the dynamical evolution of the simulated system (see the relevant sections for details on their implementation):
\begin{itemize}
\item the mass growth and planetary radius evolution of forming giant planets (Sect. \ref{sect-giant_planets})
\item orbital migration during the different phases of the formation of giant planets (Sect. \ref{sect-giant_planets})
\item the exciting effect of the self-gravity of an axisymmetric protoplanetary disk on the motion of planetary bodies (Sect. \ref{sect-protoplanetary_disk})
\item the damping effects of aerodynamic drag on the motion of the planetesimals due to the disk gas (Sect. \ref{sect-protoplanetary_disk})
\end{itemize}
For improved numerical stability and accuracy, the symplectic algorithm of {\sc Mercury-Ar$\chi$es} takes advantage of the {\sc Whfast} library \citep{rein2015} to compute the keplerian drift of the planetary bodies in place of {\sc Mercury}'s original subroutines.

\subsection{Growth and migration of the giant planets}\label{sect-giant_planets}

The growth of the giant planet in each n-body simulation follows the growth tracks from \citet{lissauer2009}, \citet{bitsch2015}, and \citet{dangelo2021} using the parametric approach from \citet{turrini2011,turrini2019}. For consistency with these works, we adopted a common formation time of 3 Myr for all giant planets in our simulations. The orbits of the giant planets lie on the midplane of the protoplanetary disk (i.e. they have inclination $i=0^\circ$) and have initial eccentricities of the order of $10^{-3}$ to account for the damping effects of the tidal gas drag on the growing planetary cores \citep[][]{cresswell2008}.

During the first $\tau_{c}=2$ Myr the giant planet accretes its core and extended atmosphere \citep{lissauer2009,dangelo2021}, its overall mass ($M_{P}$) growing from an initial value of $M_{0}=0.1\,M_{\oplus}$ (a Mars-size planetary embryo) to the critical value $M_{c}=30\,M_{\oplus}$ (equally shared between core and atmosphere) as:
\begin{equation}
 M_{P}=M_{0}+\left( \frac{e}{e-1}\right)\left(M_{c}-M_{0}\right)\left( 1-e^{-t/\tau_{c}} \right)
\label{eqn-coregrowth}
\end{equation}

After the critical mass value $M_{c}$ is reached, the giant planet enters its runaway gas accretion phase and its mass evolves over the following 1 Myr as:
\begin{equation}
 M_{P}=M_{c}+\left( M_{F} - M_{c}\right)\left( 1-e^{-(t-\tau_{c})/\tau_{g}}\right)
\label{eqn-gasgrowth}
\end{equation}
where $M_{F}$ is the final planetary mass, for which we adopted a common value of $M_{F}=1$ Jovian mass (M$_\text{J}=317.8$~M$_\oplus$). The e-folding time $\tau_{g}=10^5$ years was chosen based on the hydrodynamical simulations of \citet{lissauer2009}, \citet{coradini2010}, \citet{dangelo2010}, and \citet{dangelo2021} and references therein.

The physical radius of the growing giant planet ($R_{P}$) evolves alongside the planetary mass following the approach described by \citet{fortier2013}, which is based in turn on the hydrodynamic simulations of \citet{lissauer2009}. During the initial phase, described by Eq. \ref{eqn-coregrowth}, when the planetary core is growing its extended atmosphere, the physical radius evolves as:
\begin{equation}
 R_{P} = \frac{GM_P}{c_{s}^{2}/k_{1}+\left(GM_{P}\right)/\left(k_{2}R_{H}\right)}
 %{\frac{c_{s}^{2}}{k_{1}}+\frac{GM_P}{k_{2}R_{H}}}
 \label{eqn-inflatedradius}
\end{equation}
where $G$ is the gravitational constant, $M_P$ is the instantaneous mass of the giant planet, $c_{s}$ is the sound speed in the protoplanetary disk at the orbital distance of the planet, $R_{H}$ is the planetary Hill's radius, $k_{1}=1$ and $k_{2}=1/4$ \citep{lissauer2009}.

Once the giant planet enters its runaway gas accretion phase described by Eq. \ref{eqn-gasgrowth} (i.e. for $t > \tau_{c}$), the gravitational infall of the gas causes the planetary radius to shrink and evolve as:
\begin{equation}
 R_{P} = R_{c} - \Delta R \left(1-\exp^{-(t-\tau_{c})/\tau_{g}}\right)
 \label{eqn-collapsingradius}
\end{equation}
where $R_{c}$ is the planetary radius at $\tau_{c}=2$ Myr, i.e. the end of the extended atmosphere phase described by Eq. \ref{eqn-inflatedradius} %(or, equivalently, at the end of the core growth phase described by Eq. \ref{eqn-coregrowth})
, and $\Delta R = R_{c} - R_{I}$ is the decrease of the planetary radius during the gravitational collapse of the gas. We adopted as final value of the planetary radius $R_{I} = 1.6\,R_{J}=1.15\times10^5$ km \citep{lissauer2009, podolak2020} to account for the inflation of the young giant planet.

The migration of the growing giant planet due to its interactions with the surrounding protoplanetary disk is modelled through a semi-analytical approach that reproduces the realistic non-isothermal migration tracks from the population synthesis models by \citet{mordasini2015}. These models directly account for the interactions between a growing protoplanet and its evolving host circumstellar disk that are not included in our n-body simulations and in our parametrization of the disk environment.

As discussed in \citet{mordasini2015} and references therein, the seeds of growing protoplanets will quickly migrate (either inward or outward, depending on their position in the disk) toward the nearest convergence point, a region of zero net torque due to disk-planet interactions. Once the convergence point is reached, the protoplanet will slowly migrate inward due to Type I migration following the inward drift of the convergence point caused by the evolution of the disk \citep{mordasini2015}.

The growth of the protoplanet will eventually cause it to exit the convergence point and enter a faster regime of Type I migration. Once the protoplanet becomes massive enough to open a gap in the disk, it will transition into the slower Type II migration regime. The migration tracks of \citet{mordasini2015} show how, after the protoplanet reaches the convergence point, this sequence of different phases takes the form of a initial linear migration regime, associated with the slow migration with the convergence point, followed by a power-law migration regime, encompassing the fast Type I migration and the transition to Type II migration.

These linear and power-law regimes are reproduced in the n-body simulations using a piece-wise approach based on the analytical treatments of \citet{hahn2005} and \citet{walsh2011}. This approach allows for a computationally efficient parametrization of the dynamical behaviour of the growing giant planet and simplifies the exploration of the combined effects of planetary growth and migration.  Future studies will address the implications of physically self-consistent treatments of the coupling between disk evolution, planetary growth and migration for the planetesimal accretion history. 

During the growth of the planetary core to the critical value ($M_{C}$ in Eq. \ref{eqn-coregrowth}) the giant planet undergoes the initial slow Type I migration governed by the following drift rate $\Delta v_{1}$ \citep{hahn2005,walsh2011}:
\begin{equation}
    \Delta v_{1} = \frac{1}{2}\frac{\Delta a_{1}}{a_{p}}\frac{\Delta t}{\tau_{c}}v_{P}
\label{eqn-typeImigration}
\end{equation}
where $\Delta a_{1}$ is the radial displacement of the giant planet during its core growth, $v_{p}$ and $a_{p}$ are its current orbital velocity and semimajor axis respectively, $\tau_{c}=2$ Myr is the characteristic time governing the mass growth of the core used in Eq. \ref{eqn-coregrowth} and $\Delta t$ is the timestep used in the n--body simulations.

Once the core reaches its critical mass and the phase of runaway gas accretion begins, the giant planet first enters the faster Type I migration to later transition to the slower Type II migration. During this phase its orbital evolution is governed by the following drift rate $\Delta v_{2}$ \citep{hahn2005}:
\begin{equation}
    \Delta v_{2} = \frac{1}{2}\frac{\Delta a_{2}}{a_{p}}\frac{\Delta t}{\tau_{g}}\exp^{-\left(t-\tau_{c}\right)/\tau_{g}} v_{P}
\label{eqn-typeIImigration}
\end{equation}
where $\Delta a_{2}$ is the radial displacement of the giant planet during its runaway gas accretion,  $\tau_{g}=10^{5}$ yr is the characteristic time governing the gas accretion used in Eq. \ref{eqn-gasgrowth}, and all other parameters are the same as those in Eq. \ref{eqn-typeImigration}.

It is worth noting that the dynamical behaviour described by Eq. \ref{eqn-typeIImigration} is analogous to that adopted by \citet{shibata2020} in their simulations, where however they did not model the initial growth of the core and the associated migration described by Eq. \ref{eqn-typeImigration}. In addition, the characteristic timescale $\tau_{g}$ adopted in this work is the same as the one in their reference case, that allows for an easier comparison with their results in terms of planetesimal accretion.

As the characteristic migration timescales in Eqs. \ref{eqn-typeImigration} and \ref{eqn-typeIImigration} are fixed, the main free parameters characterizing the migration tracks are the displacements $\Delta a_{1}$ and $\Delta a_{2}$. We considered a total of six migration scenarios, with the core of the giant planet starting its growth at 5, 12, 19, 50, 100, and 130 au, respectively. In all simulations the giant planet ends its migration at 0.4 au. As we detail in Sect \ref{sect-planetesimal_disk}, this final position of the giant planet is chosen for reasons of computational efficiency. Still, the giant planet could actually end its migration anywhere within 0.4 au without its planetesimal accretion history being affected. 

\begin{figure}
    \centering
    \includegraphics[width=\hsize]{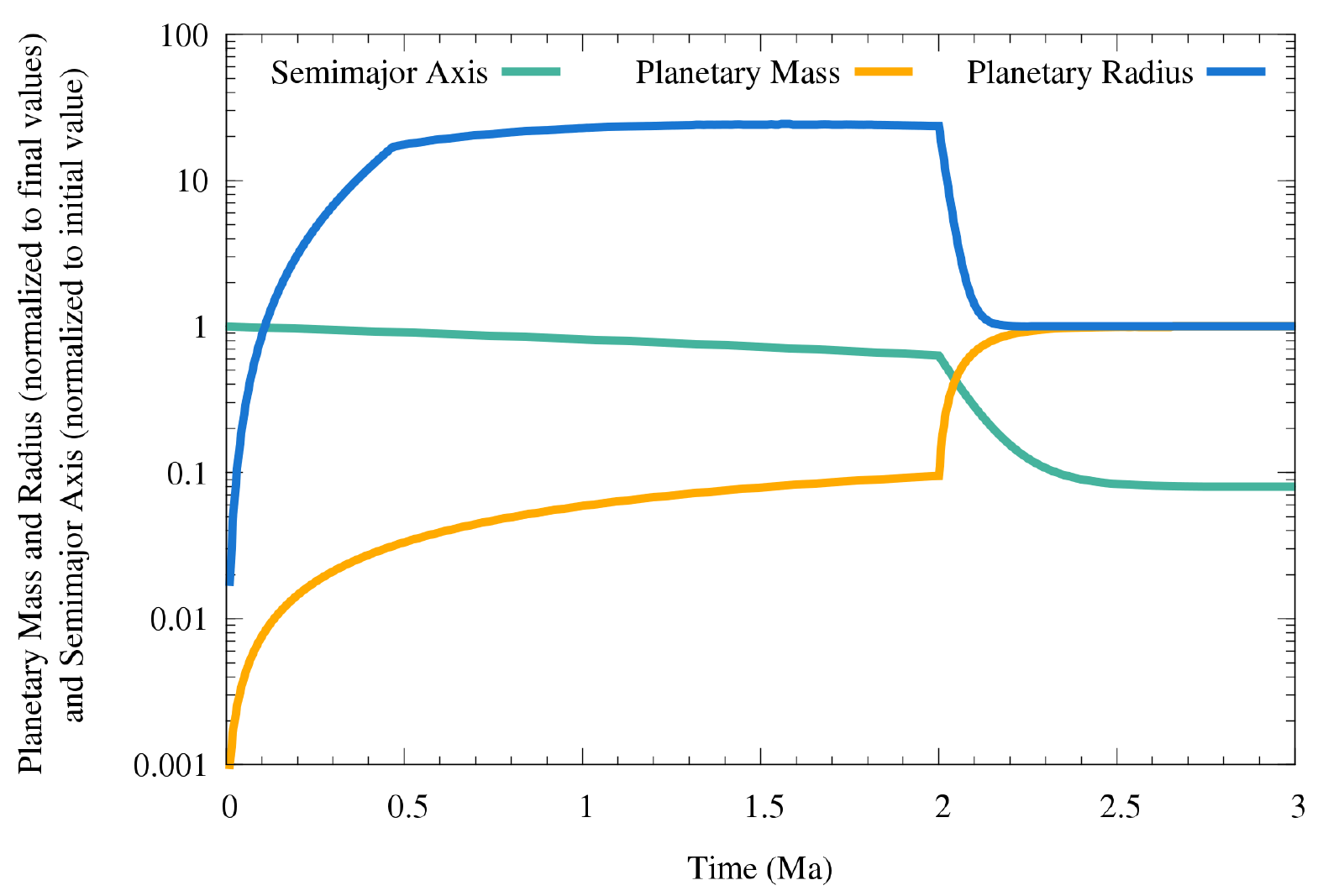}
    \caption{Normalized growth and migration tracks from the scenario where the giant planet starts forming at 5 au. The growth tracks of both mass and radius are normalized to the final mass (1~M$_{\text{J}}$) and radius (1.6~R$_{\text{J}}$) of the giant planet. The migration track is instead normalized to the initial semimajor axis (5 au) of the giant planet.}
    \label{fig-planetary_tracks}
\end{figure}

To cross-calibrate the migration tracks from \citet{mordasini2015} resulting from Eqs. \ref{eqn-typeImigration} and \ref{eqn-typeIImigration} with the joint growth and migration tracks of \citet{bitsch2015}, we defined the values of $\Delta a_{1}$ and $\Delta a_{2}$ so that they account for $40\%$ and $60\%$, respectively, of the total radial displacement the giant planet experiences during its formation in each simulation. In all scenarios the initial position of the giant planet is implicitly assumed to coincide with that of a convergence point in the disk. An example of the growth and migration tracks of the giant planet resulting from Eqs. \ref{eqn-coregrowth}-\ref{eqn-typeIImigration} is shown in Fig. \ref{fig-planetary_tracks}.

To verify that the adopted migration tracks are not associated with unrealistic migration rates, we compared them with the constraints on Jupiter's migration rates needed to reproduce the observed asymmetry among its Trojan populations \citep{pirani2019}. As discussed by \citet{pirani2019}, Jupiter's migration rate at the onset of the rapid gas accretion, when its mass was about 30\,M$_\oplus$, should have been comprised between about 30 and 300 au/Myr for the planet to capture Trojan populations consistent with current estimates.\\ 
\indent The migration rates experienced by the giant planet during the first e-folding time of the power-law migration regime of Eq. \ref{eqn-typeIImigration} range between 20 and 500 au/Myr. As a result, the scenarios where the giant planet starts between 12 and 50 au are consistent with the constraints on Jupiter's migration rate reported by \citet{pirani2019}. The scenario where the giant planet starts at 5 au is about 1/3 slower than the lower bound of the interval of values from \citet{pirani2019} while the scenarios where the giant planet starts at 100 and 130 au are about 1/3 and 2/3 faster, respectively, than the upper bound on Jupiter's migration rate.\\
\indent As a further validation, we compared the migration rate of the giant planet over the following four e-folding times (i.e. between 2.1 and 2.5 Myr), assumed as an upper limit to the Type II migration rate, with the nominal Type II migration rate associated with our parametrization of the protoplanetary disk (see Sect. \ref{sect-protoplanetary_disk}). The Type II migration rate in our scenarios thus ranges between 3 and 70 au/Myr. Following \citet{armitage2009} and adopting the same value of $\alpha=0.0053$ as \citep{bitsch2015} the nominal Type II migration rate would be of about 60 au/Myr, with the value oscillating between 1 au/Myr and 100 au/Myr for $\alpha$ comprised between $10^{-4}$ and $10^{-2}$ respectively.
%The migration tracks considered in this work are meant to explore the compositional implications of formation regions extending as far from the star as suggested by ALMA's observations for a giant planet forming over the median lifetime of about 3 Myr of protoplanetary disks \citep{fedele2010}. While the migration rates in the different scenarios are a function of the grid of initial formation distances we considered and the adopted division between the linear and exponential migration regimes discussed above. 

\subsection{Protoplanetary disks and dynamical effects of gas}\label{sect-protoplanetary_disk}

To explore the implications of wide planet-forming regions on the final compositional features of  giant planets, we modelled our template disk over one of the largest and best characterized protoplanetary disks to date, the one surrounding the 1.9 M$_{\odot}$ star HD\,163296 \citep{isella2016,turrini2019}. As the focus of our compositional modelling is on solar analogue stars (see Sect. \ref{sect-compositional_model}), we scaled down HD\,163296's observed gas surface density profile  \citep{isella2016} by a factor of 1.6 to obtain a total mass of the disk equal to the mass of a Minimum Mass Solar Nebula (MMSN, \citealt{hayashi1981}) with similar extension. 

As a result, the central star in all our simulations has a mass of 1 M$_{\odot}$ and is surrounded by a protoplanetary disk with mass M$_\text{D}=0.053$ M$_{\odot}$ whose gas surface density profile on the midplane is:
\begin{equation}
\Sigma_{gas}(r) = \Sigma_{0}\left(\frac{r}{165\,au}\right)^{-0.8} \exp{\left[-\left(\frac{r}{165\,au}\right)^{1.2}\right]}
\label{eqn-diskdensity}
\end{equation}
where $\Sigma_{0} = 3.3835$ g/cm$^{3}$. This gas surface density profile results in a disk rapidly truncating at about 500 au. We adopted a value of 0.1 au as the disk inner edge in our simulations.

The temperature radial profile in the midplane of our template disk is characterized by the same radial dependency as the disk midplane temperature estimated for HD\,163296 \citep{isella2016}. We calibrated the temperature $T_{0}$ at 1 au on the one estimated for the MMSN \citep{hayashi1981}:
\begin{equation}
T(r) = T_{0} \left(\frac{R}{1\,au}\right)^{-1/2}\,K
\label{eqn-disktemperature}
\end{equation}
where $T_{0} = 280$ K. This results in a warm protoplanetary disk where the different snow lines are distributed between $\sim$3 and $\sim$100 au (see Sect. \ref{sect-compositional_model} for details).

The density and temperature profiles of the disk from Eqs. \ref{eqn-diskdensity} and \ref{eqn-disktemperature} allow us to compute the damping effects of the gas on the dynamical evolution of the planetesimals using the treatment from \citet{brasser2007} and references therein. The drag acceleration $F_D$ is expressed by:
\begin{equation}
F_{D} = \frac{3}{8}\frac{C_{D}}{r_{p}}\frac{\rho_{g}}{\rho_{p}}v_{r}^{2}
\label{eqn-gasdrag}
\end{equation}
where $C_{D}$ is the gas drag coefficient, $\rho_{g}$ is the local density of the gas, $\rho_{p}$ and $r_{p}$ are the density and radius of the planetesimals respectively, and $v_{r}$ is the relative velocity between gas and planetesimals. 

The gas drag coefficient $C_D$ of each planetesimal is computed following the treatment described by \citet{nagasawa2019} as a function of the Reynolds ($Re$) and Mach ($Ma$) numbers, coupling the individual drag coefficients to the specific orbit of the planetesimal and to the local disk environment:
\begin{align}
C_{D} = & \left[\left(\frac{24}{Re}+\frac{40}{10+Re}\right)^{-1}+0.23Ma\right]^{-2} \nonumber \\
        & +\frac{2\cdot\left(0.8k+Ma\right)}{1.6+Ma}  
\end{align}
where $k$ is a factor with value of 0.4 for $Re < 10^{5}$ and 0.2 for $Re > 10^{5}$ \citep{nagasawa2019}.

To account for the formation of a gap in the protoplanetary disk around the growing giant planet after the onset of its runaway gas accretion phase (i.e. for $t > \tau_{c}$), we adopted the same approach as \citet{marzari2018} and assumed the giant planet to be surrounded by a gap with width:
\begin{equation}
W_{gap} = C\cdot R_{H}    
\end{equation}
where the numerical proportionality factor $C=8$ is from \citet{isella2016}. The gas density $\Sigma_{gap}(r)$ inside the gap is assumed to evolve over time with respect to the local unperturbed gas density $\Sigma(r)$ as:
\begin{equation}
\Sigma_{gap}(r) = \Sigma(r)\cdot \exp{\left[-\left(t-\tau_{c}\right)/\tau_{g}\right]}
\end{equation}
where $\tau_{c}$ and $\tau_{g}$ are the same as in Eqs. \ref{eqn-coregrowth} and \ref{eqn-gasgrowth}.

The disk density profile from Eq. \ref{eqn-diskdensity} also allows us to compute the dynamical excitation of the planetesimals due to the effects of the disk self--gravity through the analytical treatment for thin disks from \citet{ward1981}. \citet{fontana2016} showed that the perturbations due to the disk self--gravity computed with this analytical treatment are in good agreement with the perturbations computed using hydrodynamic simulations for axisymmetric disks. The effects on the self-gravity potential of non-axisymmetric density structures in the disk (e.g. \citealt{gratton2019} and reference therein for examples) are not included into this treatment and would require dedicated hydrodynamic simulations that are beyond the scope of this work. 

Here we follow the approach of \citet{nagasawa2019} and focus only on the leading term of the force due to self-gravity (F$_\text{SG}$):
\begin{equation}
F_{SG} = 2 \pi G \Sigma(r) \sum^{\infty}_{n=0} A_{n} \frac{\left(1-k\right)\left(4n+1\right)}{\left(2n+2-k\right)\left(2n-1+k\right)}
\label{eqn-selfgravity}
\end{equation}
where $k=0.8$ is the exponent of the power law part of Eq. \ref{eqn-diskdensity} and $A_{n}=\left[ (2n)!/2^{2n}(n!)^{2} \right]^{2}$ \citep{ward1981,marzari2018}. 

For this value of $k$ the sum on the right hand of Eq. \ref{eqn-selfgravity} converges to the value $-0.754126$, which means that the equation can be rewritten in more compact form as:
\begin{equation}
F_{SG} = Z \pi G \Sigma(r) 
\end{equation}
where $Z=-1.508252$. Is is shown by \citet{fontana2016} that, while the focus of \citet{nagasawa2019} on the leading perturbation term is computationally efficient, it produces accurate results only when the planetary bodies are distant from both the inner and the outer edges of the disk. For those bodies not satisfying these conditions additional terms needs to be considered in Eq. \ref{eqn-selfgravity} \citep{ward1981,fontana2016,marzari2018}.

While in our simulations the planetesimals are always well inside the outer edge of the disk (at $\approx$250 au), the same does not apply to the inner edge of the disk (at 0.1 au) at the end of the orbital migration of the giant planets. However, in the inner regions of the disk, the dynamical evolution of the planetesimals is dominated by gas drag, which quickly and efficiently cancels out any excitation effect due to the disk self-gravity. Numerical experiments have shown no appreciable difference in the dynamical evolution of the planetesimals when the additional terms required to account for the effects of the inner disk edge are included. As a result, we maintained the approach as \citet{nagasawa2019} for reasons of computational efficiency.

\subsection{Planetesimal disk and timestep of the simulations}\label{sect-planetesimal_disk}

We simulated the planetesimal disk as a swarm of dynamical tracers, each possessing an inertial mass but no gravitational mass. The lack of gravitational mass means that each tracer is affected by the gravitational potential of the star, the growing giant planet and the disk but does not affect them nor the other tracers. At the same time, each tracer possesses an inertial mass that determines the way it is affected by gas drag. 

To compute the inertial mass of the dynamical tracers we adopted a fixed radius of $r_{p}$=50 km, i.e. the characteristic size of planetesimals formed by pebble accretion \citep[e.g.][]{klahr2016}. We adopted density values of $\rho_{rock}$=2.4 g cm$^{-3}$ for the rock-dominated planetesimals inside the water snow line and $\rho_{ice}$=1 g cm$^{-3}$ for the ice-rich planetesimals beyond it (see Sect. \ref{sect-compositional_model} for details on the planetesimal composition).

The value of $\rho_{rock}$ %=2.4 g cm$^{-3}$ 
has been chosen as a compromise between the measured densities of volatile-poor and volatile-rich asteroids \citep[e.g.][and references therein]{britt2002,carry2012}. The value of $\rho_{ice}$ %=1 g cm$^{−3}$ 
has been chosen instead as a compromise between the measured densities of comets (0.4–0.6 g cm$^{-3}$; \citealt[e.g.][and references therein]{brasser2007} and \citealt{jorda2016}) and that of the larger ($\approx$200 km in diameter), ice-rich captured trans-neptunian object Phoebe (1.63 g cm$^{-3}$, \citealt{porco2005}).

The dynamical tracers representing the planetesimals are distributed randomly in semimajor axis, adopting a uniform probability distribution between 1 and 150 au (see below for the discussion of these boundaries) with a spatial density of 2000 tracers/au. The initial orbital eccentricities and inclinations (in radians) of the planetesimals are distributed randomly between 0 and 10$^{-2}$ \citep{weidenschilling2008}. 

As we do not model the growth of the core of the giant planet from the surrounding material directly, when generating the initial conditions of the planetesimal disk planetesimals are prevented from populating a radial region equivalent to the feeding zone of the giant planet to account for the local depletion of material and avoid overestimating the planetesimal accretion by the envelope.

This feeding zone is centered on the initial orbital position of the giant planet, and its width is defined through the relationship 
\begin{equation}
W_{iso} = b\cdot R_{H-iso}    
\label{eqn-gap_planetesimals}
\end{equation}%$W_{iso} = 8\cdot R_{H-iso}$ 
where $R_{H-iso}$ is the Hill's radius of a planetary core with isolation mass of $M_{iso}=15$ M$_{\oplus}$ (the critical mass used in Sect. \ref{sect-giant_planets} without the contribution of the gaseous envelope) and $b=4$ \citep[see][and references therein]{kokubo2000,dangelo2010}.

The upper boundary of the planetesimal disk at 150 au accounts for the inward drift of dust and pebbles across the formation of planetesimals due to their dynamical coupling with the gas. The inward drift of dust and pebbles results in a higher concentration of condensed materials in the inner regions of protoplanetary disks at the expense of their outer regions, which gradually deplete.

The observations of multiple circumstellar disks confirms the efficiency of this process, revealing that the gas can extend between $\approx$2 to more than 4 times farther out than the dust \citep[e.g.][]{isella2016,ansdell2018,facchini2019}. In particular, the protoplanetary disk surrounding HD\,163296, which our template disk is based on, is characterized by gas extending to about 500 au while dust is detected only up to 250 au \citep{isella2016}. 

To account for the increased concentration of solid material in the inner disk and depletion in the outer disk, we follow the approach of \citet{mordasini2009} and adopt a solids concentration factor $\xi$. The solids concentration factor $\xi$ is defined as:
\begin{equation}
\xi = \left\{
\begin{array}{ll}
     2 & \quad x \leq 150\,\rm au \\
     0 & \quad x > 150\,\rm au,
     \end{array}
    \right. 
\end{equation}
This is a conservative choice in terms of the observed range of values \citep[][]{isella2016,ansdell2018,facchini2019} that mimics the HD\,163296's radial behaviour of gas and dust.

The solids concentration factor $\xi$ is used in conjunction with the mass fraction of condensed material $Z_i$ (shown in Fig. \ref{fig-condensed_material} and described in Sect. \ref{sect-compositional_model} and Table \ref{tab-condensed_material}) to compute the radial distribution of solids
\begin{equation}\label{eqn-solids}
\Sigma_{solids}(r) = \Sigma_{gas}(r)\,\xi(r)\,Z_{i}(r)   
\end{equation}   
where the product of $\xi$ and $Z_{i}$ represents the local solids-to-gas ratio. Integrating Eq. \ref{eqn-solids} over a given annular region of the disk allows to compute the local mass of the planetesimal disk.

While our template protoplanetary disk extends down to 0.1 au (see \ref{sect-protoplanetary_disk}), for reasons of computational efficiency the initial inner boundary of the planetesimal disk is set to 1 au in the n--body simulations. This allows us to adopt a timestep $dt=15$ days (i.e. 1/24 of the orbital period of the innermost dynamical tracer) in {\sc Mercury}'s symplectic algorithm and limit the computational load of the simulations. The inclusion of planetesimals between 0.1 and 1 au would require the use of smaller timesteps, proportionally increasing the computational load. However, numerical experiments performed with planetesimal disks extending down to 0.4 au showed that tracers inside 1 au do not contribute to the accretion history of the giant planet in the growth and migration tracks we considered. 

Since planetary migration always stops at 0.4 au in our simulations (see Sect. \ref{sect-giant_planets}), near the end of its formation track the giant planet enters the orbital region where the timestep $dt=15$ days does not allow for accurately reproducing its dynamical evolution. To address this issue, we follow a similar approach to that of \citet{juric2008}. When the giant planet approaches 1 au, we decrease the timestep to $dt=2.4$ days (1/25 of the orbital period at an orbital distance of 0.3 au) to accurately resolve its orbital evolution down to its final semimajor axis. 

While the high numerical stability of the {\sc Whfast} Kepler solver \citep{rein2015} allows for this transition to occur seamlessly, for increased accuracy {\sc Mercury-Ar$\chi$es} implements a hybrid approach also during the timestep when the transition in $dt$ occurs. As a result, the keplerian orbital motion of all bodies within 1 au is computed using the Bulirsh-Stoer integrator of {\sc Mercury}  \citep{chambers1999} during this timestep.

In each simulation, we recorded the flux of impacts between the dynamical tracers representing the planetesimals and the giant planet, whose collisional cross-section matches the geometrical one associated with the physical radius computed as described in Sect. \ref{sect-giant_planets}. In principle, this treatment of the collisional cross-section might result in overestimating the flux of planetesimals impacting the giant planet. Due to their relatively large radius assumed in this work, in fact, planetesimals encountering the giant planet with high impact parameters may cross only its less dense atmospheric region and escape capture, resulting in a smaller effective collisional cross-section of the giant planet \citep[e.g.][and references therein]{fortier2013}.

It should be noted, however, that the adopted treatment neglects effects linked to the  presence and temperature of gas in a circumplanetary disk \citep[e.g.][and references therein]{coradini2010} and in the Hill's sphere of the giant planet \citep{szulagyi2016} that could increase the efficiency of the capture process. This could be achieved by enhancing the efficiency of planetesimal ablation and break-up due to the ram pressure of the gas \citep[see below and][]{mordasini2015,vazan2015, podolak2020}, by allowing for planetesimal capture in decaying circumplanetary orbits (as implicitly done in the model by \citealt{shibata2020}), or by the combined action of these effects over multiple planetary encounters \citep{podolak2020}.\\
\indent In particular, recent results by \citet{podolak2020} show that the planetesimal capture efficiency of growing giant planets quickly becomes independent on the planetesimal size as soon the mass of the gaseous envelope reaches a few Earth masses (see also below for further discussion). As a result, large planetesimals with diameter of 100 km like those considered in our simulations will be accreted with the same efficiency as small planetesimals with diameter of 1 km, for which previous studies revealed that the effective collisional radius of the giant planet closely matches its physical radius  \citep{dangelo2014,mordasini2015,vazan2015}.

We tested how planetesimal accretion scales with the collisional radius of the giant planet by simulating the scenario where the giant planet starts at 130 au while considering a collisional radius equal to 75\% the values obtained by Eqs. \ref{eqn-inflatedradius} and \ref{eqn-collapsingradius}. The results we obtained show that the planetesimal capture efficiency is about 80\% compared with the one obtained in the nominal simulation. This suggests that the role of multiple planetary encounters and of the gravitational focusing by the growing planet partly compensate for the reduced collisional cross-section and that the planetesimal accretion efficiency should scale almost linearly with the planetary radius.\\
\indent Finally, recent hydrodynamic simulations \citep{dangelo2021} report planetary radii about $\sim$25\% larger for the growing giant planet, both during the extended envelope phase (see Eq. \ref{eqn-inflatedradius}) and across the gravitational infall (see Eq. \ref{eqn-collapsingradius}), %, as well as a longer duration of the phase where the planetary core is surrounded by a massive and extended gaseous envelope, 
than those resulting from the hydrodynamic simulations of \citet{lissauer2009} considered in this work. As a result, the adopted treatment of the collisional cross-section of the giant planet should represent a reasonable and conservative approximation for the scopes of this work. We refer the readers to Sect. \ref{sect-discussion} for further discussion of this issue. 

%\textbf{A more detailed treatments of interactions between planetesimals and atmosphere  would allow for a more precise determination of the effective collisional cross-section at the cost of an increased computational load.} 

Following the approach described in \cite{turrini2011,turrini2012}, \citet{turrini2014a}, and \citet{turrini2014b}, from the knowledge of its formation region we can associate each dynamical tracer impacting on the giant planet with a mass flux of planetesimals. Specifically, integrating the solids surface density profile from Eq. \ref{eqn-solids} over a given orbital annular region of the disk allows for computing the total mass of solids, hence planetesimals, it contains. 
%integrating the gas surface density profile from Eq. \ref{eqn-diskdensity} over a given orbital \textbf{annular} region of the disk allows for computing the total mass of gas it contains. Multiplying this gas mass by the local \textbf{solids-to-gas} ratio allows for computing the total mass of solids contained in the orbital region under consideration. \textbf{As discussed above, the local \textbf{solids-to-gas} ratio is the product of the local solids concentration factor $\xi$ and mass fraction of condensed material $Z_i$.} %we refer the readers to Sect. \ref{sect-compositional_model} and Fig. \ref{fig-condensed_material} for details on both these quantities.

From the physical radius and density of the planetesimals represented by each dynamical tracer, we can compute their masses. Dividing the total mass of solids in a given orbital annular region of the disk by the individual mass of its planetesimals allows for computing the total number of planetesimals populating it. Dividing the population of planetesimals in that annular region by the number of dynamical tracers it contains, we can associate each dynamical tracer to the swarm of planetesimals it represents. From the recorded fluxes of impacting tracers, it is then straightforward to compute the mass flux of planetesimals on the giant planet. 

Recent studies \citep{dangelo2014,mordasini2015,vazan2015,podolak2020} indicate that, as the gaseous envelope of the giant planet becomes more massive than a few Earth masses, accreted planetesimals efficiently dilute into it independently on planetesimal size and composition due to the combined effects of break-up and ablation. Because of the gap we enforce around the giant planet in the intial planetesimal distribution (see Eq. \ref{eqn-gap_planetesimals}), this condition is always satisfied in our simulations before the onset of planetesimal accretion. As a result, all high-Z material accreted by the giant planet is assumed to remain mixed in the gaseous envelope \citep[see also][]{podolak2020}.

It should be noted, however, that the partitioning and distribution of high-Z material in the interiors and atmospheres of giant planets, both at the end of the formation process and after a few Gyrs of interior evolution, is still matter of debate. This uncertainty holds true even in the well-studied cases of Jupiter and Saturn in the Solar System, notwithstanding detailed \textit{in situ} measurements provided by the Galileo, Cassini and Juno missions \citep[see][and references therein for recent in-depth discussions]{atreya2018,helled2018,stevenson2020}.

The combination of the data provided by the Galileo and Juno missions for Jupiter suggest the diffuse presence of high-Z elements both in the gaseous envelope and the atmosphere of the giant planet, although little to no information is available on their possible compositional homogeneity or radial gradients \citep[see][and references therein]{stevenson2020}. In this work the high-Z material accreted through planetesimals is assumed to be uniformly distributed into the envelope, leaving to future work to address the issue in more detail. We refer the readers to Sect. \ref{sect-discussion} for a discussion of the implications of a non-uniform distribution of the high-Z material.
%The effect of convective mixing, under conditions common to giant planets, can spread initial compositional gradients (e.g. a concentration of high-Z material in the inner envelope) to the outer envelope over time \citep{vazan2015}. As a result, in our analysis, the high-Z material accreted through planetesimals is assumed to be uniformly distributed into the envelope.} 

%We refer interested reader to \citet{turrini2011,turrini2012,turrini2014a,turrini2014b} for further details on this approach.

\subsection{Compositional model of the disk and the planetesimals}\label{sect-compositional_model}

\begin{table}[t]
    \centering
    \caption{Protosolar and meteoritic abundances of C, O, N and S}    \label{tab-protosolar_vs_meteoritic}
    \begin{tabular}{c c c c }
    \hline
    Element & Protosolar   & Meteoric   & Protosolar/     \\
            & Abundance    & Abundance  & Meteoritic Ratio  \\
    \hline
    C   &  $3.0\times10^{-4}$  &  $2.6\times10^{-5}$  &  0.09  \\
    O   &  $5.4\times10^{-4}$  &  $2.6\times10^{-4}$  &  0.48  \\
    N   &  $7.4\times10^{-5}$  &  $1.9\times10^{-6}$  &  0.03  \\
    S   &  $1.5\times10^{-5}$  &  $1.5\times10^{-5}$  &  1.00  \\
    \hline
    \end{tabular}
    %\caption{Estimated protosolar and meteoritic abundances of C, O, N and S atoms expressed as relative abundances with respect to H atoms%$10^{12}$ H atoms %(i.e. the astronomical abundance scale, \citealt{asplund2009,lodders2010}).}
    %.}
    \justify
    Note. Estimated abundances are expressed as relative abundances with respect to 10$^{12}$ H atoms \citep{asplund2009,lodders2010}.
\end{table}

\begin{table*}[t]
    \centering
    \caption{Protostellar abundances and partitioning of C, O, N, S among rocks, organics and ices}\label{tab-elemental_abundances}
    \begin{tabular}{c c c c c}
    \hline
    Element & Protostellar & \multicolumn{3}{c}{Partition among phases} \\
    \cline{3-5}
            & Abundance    & Rock       &  Refractory Organics     &  Ice   \\
    \hline
    C   &  $3.0\times10^{-4}$  &  $2.6\times10^{-5}$  & $1.9\times10^{-4}$ & $8.0\times10^{-5}$\\
    O   &  $5.4\times10^{-4}$  &  $2.6\times10^{-4}$  & 0  &  $2.8\times10^{-4}$ \\
    N   &  $7.4\times10^{-5}$  &  $1.9\times10^{-6}$  & 0  &  $7.2\times10^{-5}$   \\
    S   &  $1.5\times10^{-5}$  &  $1.5\times10^{-5}$  & 0  &  0  \\
    \hline
    \end{tabular}
    \justify
    \textbf{Note. }Protostellar abundances of C, O, N and S in our compositional model expressed as relative abundances with respect to H %in the astrophysical scale (number of atoms every $10^{12}$ H atoms) 
    and their partitioning in condensed form between rocks, organics and ices. We refer the readers to the main text for more details on the different solid phases and the adopted abundances.
\end{table*}

\begin{figure}
    \centering
    \includegraphics[width=\hsize]{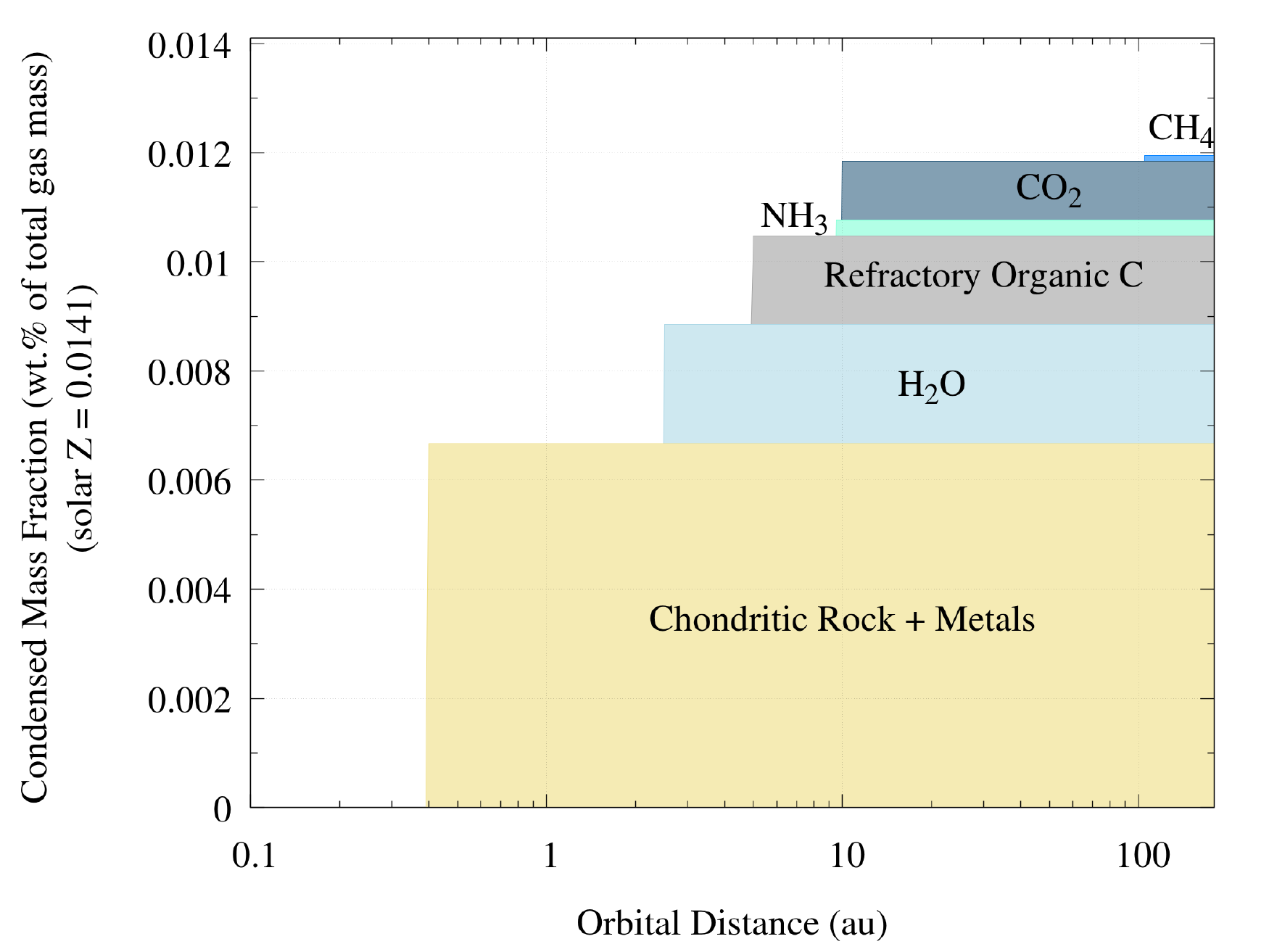}
    \caption{Mass fraction of solid material as a function of the distance from the star in our template protoplanetary disk. The mass fraction is expressed with respect to the total mass of gas assuming a solar composition for the latter. The mass fraction of condensed material is always lower than the protostellar Z as some elements (Ne) and molecules (CO and N$_2$) never condense in our disk (see main text and Table \ref{tab-condensed_material} for further details).}
    \label{fig-condensed_material}
\end{figure}

\begin{figure}
    \centering
    \includegraphics[width=\hsize]{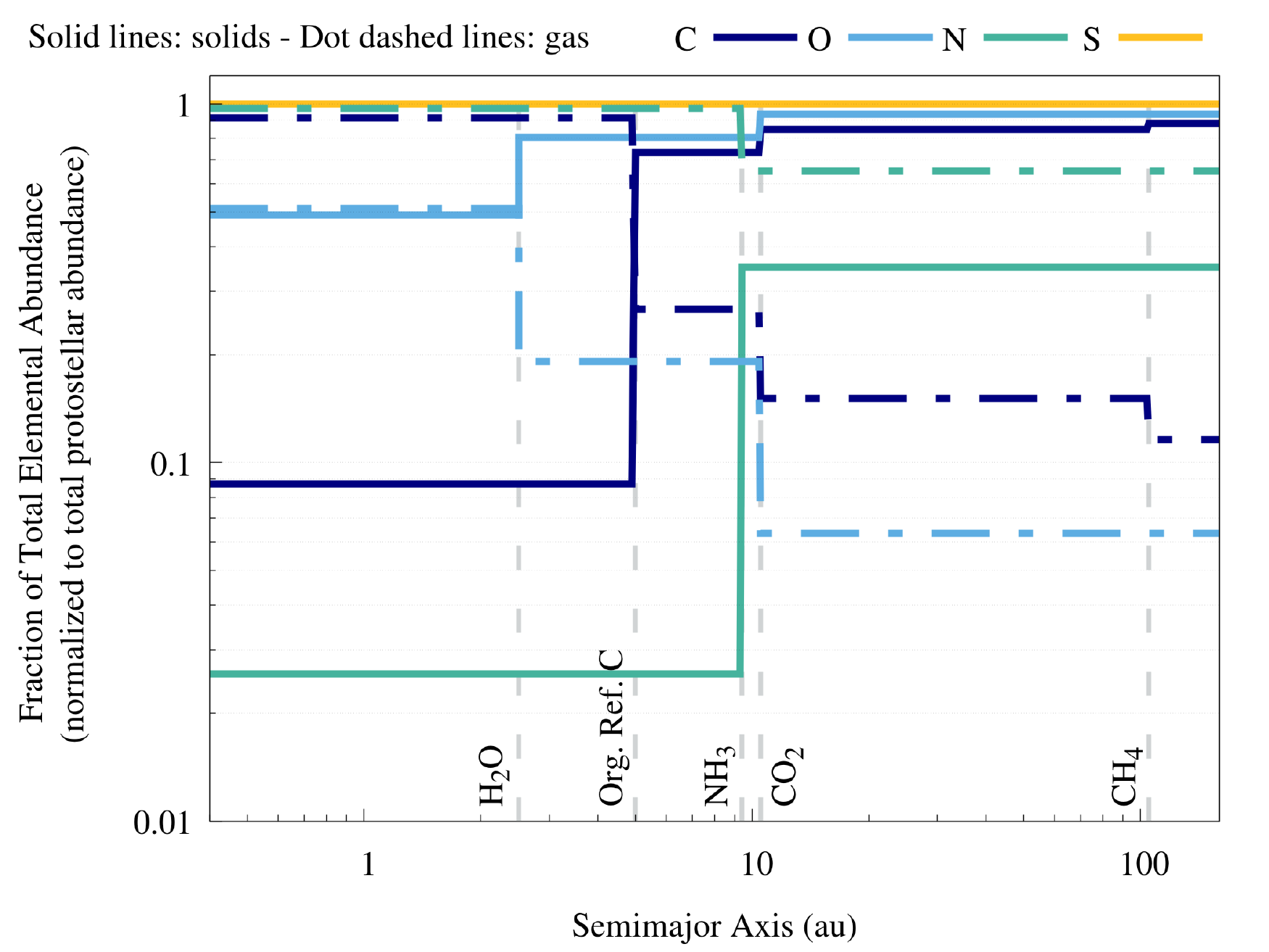}
    \caption{Fractions of C, O, N and S in solid phase (solid lines) and gas phase (dot dashed lines) in the protoplanetary disk at varying distances from the host star. The values are normalized to the protostellar abundances reported in Tables \ref{tab-protosolar_vs_meteoritic} and \ref{tab-elemental_abundances}. The vertical dashed lines indicate the position of the snowlines of H$_2$O (2.5 au), refractory organic carbon (5 au), NH$_3$ (9.4 au), CO$_2$ (10.5 au) and CH$_4$ (105 au).}
    \label{fig-element_partitioning}
\end{figure}

\begin{table*}[t]
    \centering
    \caption{Condensation sequence and total mass fraction $Z_{i}$ in solid phase across the protoplanetary disk}   \label{tab-condensed_material}
    \begin{tabular}{c c c c c c c}
    \hline
    %Material/Molecule  
    & Rocks    & H$_{2}$O & Ref. Org. C & NH$_{3}$ & CO$_{2}$ & CH$_{4}$ \\
    \hline
    Snow Line (au) & - & 2.5 & 5.0 & 9.5 & 10.5 & 105 \\ 
    $Z_{i}$ & $6.6\times10^{-3}$ & $8.8\times10^{-3}$ & $1.04\times10^{-2}$ & $1.07\times10^{-2}$ & $1.18\times10^{-2}$ & $1.19\times10^{-2}$ \\
    \hline
    \end{tabular}
    %\caption{Estimated protosolar and meteoritic abundances of C, O, N and S atoms expressed as relative abundances with respect to H atoms%$10^{12}$ H atoms %(i.e. the astronomical abundance scale, \citealt{asplund2009,lodders2010}).}
    %.}
    \justify
    \textbf{Note.} Cumulative mass fraction $Z_{i}$ of all materials/molecules in solid phase across the different compositional regions of the protoplanetary disk due to the condensation of rocks, organics and ices as illustrated in Fig. \ref{fig-condensed_material}. For each material/molecule in the condensation sequence except rocks we report the position of the associated snowline. Each value of $Z_{i}$ represents the total mass fraction resulting from the whole condensation sequence up to the relevant snowline (e.g. the value at 9.5 au is due to the sum of rocks, H$_2$O, refractory organic C and NH$_3$).
\end{table*}

As introduced in Sect. \ref{sect-protoplanetary_disk}, we focus our study on host stars that are solar analogues both in terms of mass (1 M$_\odot$) and composition. The protostellar elemental abundances of both star and disk are computed from the photospheric abundances from \citet{asplund2009} integrated with the updated values from \citet{scott2015a,scott2015b} and corrected for the sinking of He and the high-Z elements as described in \citet{asplund2009}. The values X=0.7148, Y=0.2711 and Z=0.0141 characterize the resulting composition of both star and disk in terms of mass fractions of H, He and heavy elements, respectively.

Our compositional model focuses on four tracing elements: carbon (C), oxygen (O), nitrogen (N), and sulphur (S). To characterise their distribution across the different phases (gas, rock, organics, ice) in the protoplanetary disk we follow an approach conceptually similar to previous works \citep[e.g.][]{johnson2012,thiabaud2014,marboeuf2014a,marboeuf2014b,mordasini2016,cridland2019} and combine the information provided by meteoritic abundances \citep{lodders2010,palme2014}, extrasolar planetary material \citep{doyle2019,kama2019}, the interstellar medium \citep{palme2014,bergin2015,mordasini2016} and astrochemical models \citep{eistrup2016,eistrup2018}.

The first step in our compositional modelling is to determine the mass fraction of gas condensed as rock in the inner regions of the protoplanetary disk. In the framework of our study, it is not important whether this mass fraction initially condenses as dust, pebbles or planetesimals, as long as planetesimals incorporate the bulk of it by the time the giant planet starts migrating. Comparing solar and meteoritic abundances and assuming that the rock-forming elements are condensed in meteoritic proportion \citep{lodders2010,palme2014}, the mass fraction of disk gas condensing as rock is $Z_{rock}= 6.6\times10^{-3}$ (see Fig. \ref{fig-condensed_material} and Table \ref{tab-condensed_material}). 

By subtraction from the protostellar Z, the mass fraction of high-Z material that remains in the gas phase as volatiles, progressively to condense at larger distances as ices and organics, amounts to $Z_{ice}=7.5\times10^{-3}$. About 17\% of $Z_{ice}$ is provided by Ne, an element not included among our tracing elements but nevertheless contributing a significant fraction of the total high-Z material (see \citealt{asplund2009,lodders2010,palme2014}). Moreover, some molecules of the four elements we consider (specifically, CO and N$_{2}$) never condense in the protoplanetary disk due to the adopted thermal profile. This means that the condensed fraction of high-Z elements across our template disk is always lower than the protostellar Z value%, with $Z_{ice}=6.4\times10^{-3}$ representing its maximum value when all C, O and N condense. 
, with $Z_{ice}=5.3\times10^{-3}$ representing the maximum mass fraction of condensed volatiles in the outer disk (see below,  Fig. \ref{fig-condensed_material} and Table \ref{tab-condensed_material}).

The next steps involve distributing the four tracing elements across their carriers while preserving the mass balance. The protostellar and meteoritic abundances for the four tracing elements are reported in Table \ref{tab-protosolar_vs_meteoritic}, together with their protostellar/meteoritic ratio \citep[see][for details on the comparison methodology]{asplund2009,lodders2010,palme2014}. The distribution of the four tracing elements across the gas and solid phases is shown in Fig. \ref{fig-element_partitioning} while their partition among the different carriers (rocks, organics, ices) is reported in Table \ref{tab-elemental_abundances}. 

Based on meteoritic data for CI carbonaceous chondrites \citep{lodders2010,palme2014}, chondritic rocky material carries the totality of S and about half the O, while only 9$\%$ of C and 3$\%$ of N is contained in rocks (see Fig. \ref{fig-element_partitioning} and Table \ref{tab-elemental_abundances}). It is worth highlighting that chondritic rocks contain significantly more O than would be expected from a pure mixture of silicon, magnesium and iron oxides (which would account only for $\sim$27\% of the total O, \citealt{lodders2010,palme2014}), likely due to the presence of hydrated minerals (e.g. phyllosilicates) or refractory organic material \citep{pollack1994,semenov2003}.

%We also want to highlight that in Table \ref{tab-protosolar_vs_meteoritic} the meteoritic abundance of S is slightly larger (by about $2.3\%$) than the protosolar value: this discrepancy, due to the different measurement techniques, falls within the accepted uncertainties in such comparisons ($5-10\%$, \citealt[see][and references therein]{lodders2010,palme2014,palme2017}). As shown in Table \ref{tab-elemental_abundances}, we adopted the protosolar value of S as our reference value.

The fractional abundance (48\%) of O carried in refractory form in rocks determined above is consistent with the data on O fugacities of exoplanetary materials gathered from observations of polluted white dwarfs \citep{doyle2019}. In parallel, the fractional abundance (100\%) of S carried in refractory form in rocks determined from meteoritic data is consistent with the observations of the composition of material accreted by stars in a sample of young disk-hosting B-F stars \citep{kama2019}. The comparison with the extrasolar data also reveals that the differences between the refractory/rocky component in our model, based on the protosolar composition, and analogous exoplanetary materials are of the order of $\sim$10\%.

The observations of the composition of comet 67P Churyumov-Gerasimenko (67P C-G in the following) by the ESA mission Rosetta provide further support for the leading role of refractory materials as the main carrier of S, in agreement with the results of \citet{kama2019}. Rosetta's observations show a relative abundance of $\rm H_{2}S$, the main volatile carrier of S, of the order of $10^{-2}$ that of water \citep[][]{leroy2015,rubin2019,rubin2020}. In our compositional model this amounts to $\sim$10\% of the solar budget of S (see below, and Table \ref{tab-elemental_abundances} and Fig. \ref{fig-element_partitioning}). %for the abundance of oxygen in volatile form as water). 
As a result, both cometary and extrasolar data constrain the accuracy of our partitioning of S % in terms of refractory S 
to about 10\%. %Furthermore, it is important to note that, since our model focuses on elemental abundances, it is not influenced by the specific form in which S is carried (refractory or volatile), only by its total abundance.}

The remaining fraction (52\%) of O not incorporated in rocks and the majority of C and N are in the form of gas in the inner regions of the protoplanetary disks (see Fig. \ref{fig-element_partitioning}). We took advantage of the astrochemical models from \citet{eistrup2016,eistrup2018} to determine how they partition across the different molecules and the different phases. Specifically, we used the abundances (rescaled as discussed below to preserve mass balance) and condensation profiles of the different volatile molecules to estimate the abundances of C, O, and N atoms in the gas and condensed as ices as a function of the distance from the host star. Using the relevant molecular weights, we computed the mass contribution of each molecule of these elements to the condensed mass fraction $Z_{i}$ (see Fig. \ref{fig-condensed_material} and Table \ref{tab-condensed_material}).

Among the evolutionary scenarios investigated by \citet{eistrup2016,eistrup2018} we focused on the one where the abundances of ices in planetesimals are inherited from the pre-stellar phase (labelled ``inheritance - low ionization'' scenario in \citealt{eistrup2018}). This scenario was chosen based on the results of the comparison between Rosetta's observations of comet 67P C-G and the protostellar systems IRAS 16293-2422 and SVS13-A \citep{drozdovskaya2019,bianchi2019}, which suggest a significant imprint of interstellar chemical abundances on cometary ices \citep[see also][for a discussion]{altwegg2019}. \\
\indent Our reference for the partitioning of C, O, and N among the different molecules is based on the results of the astrochemical models of \citet{eistrup2018} after 1 Myr of evolution of the protoplanetary disk, for reasons we will discuss in the following. Before proceeding, it should be noted that the adopted disk chemical scenario is not unique: a discussion of the implications of different chemical scenarios for our analysis is provided in Sect. \ref{sect-caveats}.

As shown in Fig. \ref{fig-element_partitioning}, in the inheritance scenario from \citet{eistrup2018} the bulk abundance of N is partitioned between NH$_{3}$ ($\sim$33$\%$) and N$_{2}$ ($\sim$66$\%$). Since the total abundance of N with respect to hydrogen ([N/H]) reported by \citet{eistrup2016} ($6.2\times10^{-5}$) is lower than the protostellar value derived from \citet{asplund2009} and reported in Table \ref{tab-protosolar_vs_meteoritic}, we scaled the total abundances of N, NH$_{3}$ and N$_{2}$ from \citet{eistrup2016,eistrup2018} by a common factor of 1.16. This ensures the conservation of the total mass of N in the protostellar disk, including the fraction incorporated in rocks (see Table \ref{tab-elemental_abundances}), while preserving the chemical network and results of \citet{eistrup2016,eistrup2018}.

In the case of O, the abundance of $5.2\times10^{-4}$ reported for ices by \citet{eistrup2016,eistrup2018} is close to the protostellar abundance derived from \citet{asplund2009} and reported in Table \ref{tab-protosolar_vs_meteoritic}. This, however, neglects the amount of O sequestered in refractory form into chondritic rocks (see Fig. \ref{fig-element_partitioning} and Tables \ref{tab-protosolar_vs_meteoritic} and \ref{tab-elemental_abundances}). We therefore scaled the abundance of O from \citet{eistrup2016,eistrup2018} by a factor of 0.54 to match the fraction of O in volatile form in our protoplanetary disk (see Table \ref{tab-elemental_abundances}). Since O and C share a joint chemical network, we scaled the abundance of C of $1.8\times10^{-4}$ from \citet{eistrup2016,eistrup2018} by the same factor of 0.54 to preserve the relative proportions of C- and O-bearing volatiles of their astrochemical models (see Table \ref{tab-elemental_abundances}). 

This scaling preserves the mass balance of O throughout the disk but leaves a fraction of C unaccounted for. However, previous work comparing the C/Si values observed in meteorites with those of interstellar dust \citep[see][and references therein]{bergin2015} highlighted the significant role played by carbonaceous dust and/or C-based organic compounds as reservoirs of solid C in planetary materials. The contribution of such reservoirs can account for $\sim$50\% of the total carbon budget, exceeding that of rocks and C-based ices \citep{bergin2015}, as recently supported by Rosetta's measurements of the C-rich composition of 67P C-G's dust \citep{bardyn2017,isnard2019}.

We followed \citet{cridland2019} and previous authors \citep[e.g.][]{mordasini2016,bergin2015,thiabaud2014} and included this additional C reservoir in our compositional model alongside rocks and ices. In particular, following \citet{cridland2019} we increased the amount of C in solid form included in planetesimals from 5 au outward to match the C/Si=6 value observed in comet 67P C-G's dust \citep{bardyn2017} and in the interstellar medium (see Figs. \ref{fig-condensed_material} and \ref{fig-element_partitioning}, and Table \ref{tab-condensed_material}). This allows us to account for all missing C after the scaling described above. The transition at 5 au accounts for the destruction of carbonaceous dust in the inner regions of protoplanetary disks \citep{bergin2015}. To preserve the C mass balance throughout the disk, the gas in the inner 5 au is proportionally enriched in C in the form of %assumed to be in reduced form 
C-H bond molecules, as suggested by Rosetta's observations of the high H/C ($\approx$1) ratio in 67P C-G's dust. 

We adopt the naming convention from \citet{thiabaud2014} and \citep{isnard2019} and refer to this C reservoir as refractory organic C to distinguish it from the C included in chondritic rocks (see Table \ref{tab-elemental_abundances}). Before proceeding, it is important to note that the focus of our compositional model is on the distribution of the four tracing elements across the gas and solid phases in the disk (see Fig. \ref{fig-element_partitioning}), not on the specific form in which the elements are incorporated into the planetesimals. Specifically, our model is not influenced by whether the additional C is incorporated as graphite, nano-diamonds or C-H bond chains (a similar point is valid also for the incorporation of O in phyllosilicates or refractory organics). 

From the abundances of the different volatile and organic carriers of C, O, N discussed above and the results of the astrochemical models of \citet{eistrup2016,eistrup2018}, we can compute how the mass fraction of condensed material grows throughout our template protoplanetary disk: the result and the associated compositional gradient are shown Fig. \ref{fig-condensed_material} and Table \ref{tab-condensed_material}. As illustrated in Figs. \ref{fig-condensed_material} and \ref{fig-element_partitioning} and reported in Table \ref{tab-condensed_material}, the snow lines of H$_2$O, NH$_3$, CO$_2$, and CH$_4$ are located at 2.5, 9.5, 10.5 and 105 au, respectively. Note that the total mass fraction of condensed material is always lower than the protostellar Z value due to the missing contribution of CO and N$_{2}$, since they never condense in the thermal environment of our template disk, alongside that of Ne as discussed before.

It should be noted that our disk model assumes both thermal and compositional structures that are constant in time (see Eq. \ref{eqn-disktemperature} and Figs. \ref{fig-condensed_material} and \ref{fig-element_partitioning}). In evolving protoplanetary disks, however, the positions of the snowlines will drift inward over time in response to the evolving disk environment and will affect its compositional structure \citep[][and references therein]{eistrup2018}.

%\textbf{The largest changes in the position of the snowlines ($\sim$10-20\%) are expected over the first 1 Myr of life of circumstellar disks, while changes over each subsequent 1 Myr-wide temporal interval become progressively less marked \citep{eistrup2018}. In our disk model, the snowline most visibly affected by this drift would be that of CH$_{4}$ (see also \citealt{eistrup2018}), which however contributes in a limited way to both the solid and C mass fractions (see Figs. \ref{fig-condensed_material} and \ref{fig-element_partitioning}).}

Recent comparisons between the masses of exoplanetary systems and of the dust and gas in protoplanetary disks suggest that the bulk of the planetesimal formation process should occur rapidly on a timescale of 1 Myr over the whole disk \citep{manara2018}, consistently with meteoritic data from the Solar Sytem \citep[e.g.][]{scott2007}. The conversion of dust into planetesimals should cause gas-grain chemistry to become increasingly less efficient and the chemical evolution of the disk to slow down with respect to what would be expected in a dust-rich environment.

%\textbf{In parallel, the bulk of the gas accretion by the giant planet occurs during the rapid gas accretion phase \citep[e.g.][and references therein]{lissauer2009,dangelo2020}, which lasts a fraction of Myr \citep[see Eq. \ref{eqn-gasgrowth} and][]{lissauer2009,bitsch2015,dangelo2020,shibata2020}. Also in this case, the compositional structure of the protoplanetary disk can be reasonably approximated as constant during the process.}

As a consequence, once most of the mass initially present as dust is converted into planetesimals, the compositional structure of the protoplanetary and planetesimal disks can be reasonably approximated as fixed. As discussed above, consistently with the timescale reported by \citet{manara2018} we adopted the results of the astrochemical models of \citet{eistrup2018} after 1 Myr of evolution of the protoplanetary disk as our reference. %The underlying assumption of thermal and compositional structures constant in time of our disk model can therefore be considered a reasonable approximation. 
The implications of temporally-evolving disk models will be considered in future works. We refer the readers to Sect. \ref{sect-discussion} for further discussion of this issue.

\section{Results}\label{sect-results}

\begin{table*}[tb]
    \centering
    \caption{Formation scenarios, envelope metallicity and compositions of the giant planets}\label{tab-simulation_results}
    \begin{tabular}{c c c c c c c c c}
    \hline
    Formation & Initial Core & \multicolumn{3}{c}{High-Z Elements Accreted}  & Total C & Total O & Total N & Total S \\
    \cline{3-5}
    Scenario & Position (au)             & Solids (M$_{\oplus}$) & Gas (M$_{\oplus}$) & Total (M$_{\oplus}$) &  &  &  &  \\
    \hline
    1	&	5	&	2.2	&	2.1	&	4.3	&	$2.8\times10^{-4}$	&	$5.7\times10^{-4}$	&	$7.6\times10^{-5}$	&	$1.2\times10^{-5}$	\\
    2	&	12	&	3.9	&	1.6	&	5.5	&	$4.0\times10^{-4}$	&	$6.9\times10^{-4}$	&	$8.1\times10^{-5}$	&	$1.8\times10^{-5}$	\\
    3	&	19	&	6.3	&	1.4	&	7.7	&	$5.3\times10^{-4}$	&	$1.0\times10^{-3}$	&	$1.0\times10^{-4}$	&	$2.7\times10^{-5}$	\\
    4	&	50	&	13.6	&	1	&	14.6	&	$9.7\times10^{-4}$	&	$2.0\times10^{-3}$	&	$1.4\times10^{-4}$	&	$5.6\times10^{-5}$	\\
    5	&	100	&	19.5	&	0.9	&	20.4	&	$1.4\times10^{-3}$	&	$2.8\times10^{-3}$	&	$1.8\times10^{-4}$	&	$8.0\times10^{-5}$	\\
    6	&	130	&	31.8	&	0.8	&	32.6	&	$2.2\times10^{-3}$	&	$4.5\times10^{-3}$	&	$2.7\times10^{-4}$	&	$1.3\times10^{-4}$	\\
    \hline
    \multicolumn{2}{c}{Reference values (solar mixture)} & - & - & 4.3 & $3.0\times10^{-4}$ & $5.4\times10^{-4}$ & $7.4\times10^{-5}$ & $1.5\times10^{-5}$ \\
    \hline
    \end{tabular}
    \justify
    \textbf{Note.} Formation scenarios considered in this work and resulting composition of the gaseous envelope of the giant planet. For each scenario we report the initial position of the giant planet, the accreted mass of high-Z elements due to planetesimals and gas as well as the total one due to both contributions, and the final abundances with respect to H of C, O, N and S in the gaseous envelope. For reference, at the bottom of the table we report also the values the gaseous envelope would possess if it were characterized by a purely solar composition. Note how in scenario 1 the quantity of high-Z elements matches the solar one yet O and N are super-solar while C and S are sub-solar.
\end{table*}

\begin{table}%[tb]
    \centering
    \caption{Gas contributions to the C, O and N abundances of the gaseous envelope}\label{tab-simulation_gas_only}
    \begin{tabular}{c c c c}
    \hline
    Formation   & \multicolumn{3}{c}{Gas Contribution} \\
    \cline{2-4}
    Scenario    &   C &   O &   N   \\
    \hline
    1	&	$2.5\times10^{-4}$	&	$2.3\times10^{-4}$	&	$7.4\times10^{-5}$	\\
    2	&	$1.8\times10^{-4}$	&	$1.5\times10^{-4}$	&	$7.1\times10^{-5}$	\\
    3	&	$1.4\times10^{-4}$	&	$1.2\times10^{-4}$	&	$6.6\times10^{-5}$	\\
    4	&	$7.6\times10^{-5}$	&	$6.5\times10^{-5}$	&	$5.5\times10^{-5}$	\\
    5	&	$5.7\times10^{-5}$	&	$4.7\times10^{-5}$	&	$5.2\times10^{-5}$	\\
    6	&	$5.3\times10^{-5}$	&	$4.3\times10^{-5}$	&	$5.1\times10^{-5}$	\\
    \hline
    Solar & $3.0\times10^{-4}$ & $5.4\times10^{-4}$ & $7.4\times10^{-5}$ \\
    \hline
    \end{tabular}
    \justify
    \textbf{Note.} Gas contributions to the C, O and N abundances in the gaseous envelopes of the giant planet for the scenarios reported in Table \ref{tab-simulation_results}. At the bottom of the table we also report the respective contributions of an equivalent amount of gas with solar composition.
\end{table}

As introduced in Sect. \ref{sect-model}, we investigated a set of six formation and migration scenarios with the seeds of the giant planet starting their growth and migration at different distances from their host star. As summarized in Table \ref{tab-simulation_results}, the initial semimajor axes of the giant planet we considered were 5, 12, 19, 50, 100 and 130 au respectively, with the giant planet concluding its migration at 0.4 au in all scenarios. In each simulation, we recorded the flux of impacts between the dynamical tracers representing the planetesimals and the giant planets, from which we then computed the total accreted mass of solid material as detailed in Sect. \ref{sect-planetesimal_disk}. 

Using the compositional model described in Sect. \ref{sect-compositional_model} we then converted the accreted masses of gas and planetesimals into the total mass of accreted high-Z elements and the atomic abundances of C, O, N and S with respect to H in the gaseous envelope of the giant planet. As discussed in Sect. \ref{sect-planetesimal_disk}, high-Z elements are assumed to be uniformly distributed in the gaseous envelope. The outcomes of our simulations are shown in Tables \ref{tab-simulation_results} and \ref{tab-simulation_gas_only}. Since our compositional model focuses on host stars that are solar analogues, in the following we will refer to solar abundances and stellar abundances interchangeably.

\subsection{Planetesimal accretion, orbital migration and planetary metallicity}\label{sect-enrichment}

A first, immediate result shown by Table \ref{tab-simulation_results} is the direct correlation between the extent of migration and the enrichment of the giant planet due to the accretion of planetesimals, that is, the largest the migration, the greatest the enrichment. This result is in line with the findings of the work by \cite{shibata2020}, where they investigated the enrichment in heavy elements by migrating giant planets in disks more compact and more massive than the one we considered in our work. The range of total envelope metallicities reported in Table \ref{tab-simulation_results} spans the whole 1-$\sigma$ uncertainty range in the mass-metallicity trend of Jovian planets from \citet{wakeford2017} and \citet{sing2018} based on Solar System and exoplanetary data (see also Table \ref{tab-elemental_ratios}).

As can be seen in Table \ref{tab-simulation_results}, the metallicity of the gaseous envelope is well approximated by the accreted mass of planetesimals when giant planets undergo extensive migration (scenarios 4-6). However, in case of moderate migration, the two quantities become increasingly different due to the growing contribution from the accreted gas (scenarios 1-3, see also Table \ref{tab-simulation_gas_only}). The transition between the two regimes will depend on the extension and  metallicity of the host protoplanetary disk. Extrapolating from the trend in Table \ref{tab-simulation_results}, in case of limited migration the final metallicity of the planetary envelope can be sub-stellar, in line with the findings by \cite{turrini2015} and references therein,  \cite{shibata2019}, and \cite{shibata2020}. 

Once considered together, our results and those of \cite{shibata2020} support the view that giant planets characterized by super-stellar metallicity values should have formed at markedly larger distances from their host stars than their counterparts characterized by sub-stellar metallicity values, unless the latter formed by a different mechanism (e.g. by disk instability). In the case of large protoplanetary disks like that considered in this work, our results indicate that super-stellar metallicity values require formation regions consistent with those of the giant planets indirectly identified by ALMA's observational surveys \citep[e.g.][]{alma2015,isella2016,fedele2017,fedele2018,long2018,andrews2018}.

When the atomic abundances of C, O, N, and S with respect to H are computed from the accreted masses of gas and planetesimals, the results reported in Table \ref{tab-simulation_results} also show how the final planetary composition is markedly non-solar/non-stellar and highlight the lack of simple and common relationships between metallicity and atomic elemental abundances. An immediate example is provided by scenario 1 where the giant planet starts to form at 5 au: the metallicity is solar, but O and N are super-solar while C and S are sub-solar.

\begin{table*}%[tb]
    \centering
    \caption{Elemental ratios and metallicity of the gaseous envelope}\label{tab-elemental_ratios}
    \begin{tabular}{c c c c c c c c c c c c}
    \hline
    Formation   & & \multicolumn{5}{c}{Gas + Solids} & & \multicolumn{3}{c}{Gas Only}\\
    \cline{3-7} \cline{9-12}
    Scenario    &  & Z &  C/O & N/O & C/N & S/N &  & Z & C/O & N/O & C/N \\
    \hline
    1	& &	1.0 & 0.49 & 0.13	&	3.64	&   0.16 & & 0.5 & 1.13	&	0.33	&	3.43  \\
    2	& &	1.3 & 0.58 & 0.12	&	4.95	&	0.22 & & 0.4 & 1.24	&	0.48	&	2.57  \\
    3	& & 1.8 & 0.53 & 0.10	&	5.31	& 	0.27 & & 0.3 & 1.15	&	0.56	&	2.07  \\
    4	& &	3.4 & 0.50 & 0.07	&	6.80	& 	0.39 & & 0.2 & 1.18	&	0.86	&	1.38  \\
    5	& &	4.8 & 0.50 & 0.07	&	7.54	& 	0.44 & & 0.2 & 1.22	&	1.10	&	1.11  \\
    6	& &	7.6 & 0.50 & 0.06	&	8.28	& 	0.48 & & 0.2 & 1.22	&	1.18	&	1.04  \\
    \hline
    Solar & & 1.0 & 0.55 & 0.14 & 3.98 & 0.20 & & 1.0 & 0.55 & 0.14 & 3.98 \\
    \hline
    \end{tabular}
    \justify
    \textbf{Note.} Elemental ratios and metallicity Z (in units of solar metallicity) of the gaseous envelopes of the giant planet in the different formation scenarios when considering the contributions of both gas and solids and of gas only. At the bottom of the table we report the respective solar elemental ratios for comparison. The metallicity Z in the gas and solids case is computed from the total accreted high-Z material in Table \ref{tab-elemental_abundances}, while in the gas only case it is computed from the gas-accreted high-Z material reported in the same table.
\end{table*}

\subsection{C/O ratios of gas-dominated and solid-enriched giant planets}\label{sect-C_O_ratio}

\begin{figure}[ht]
    \centering
    \includegraphics[width=\hsize]{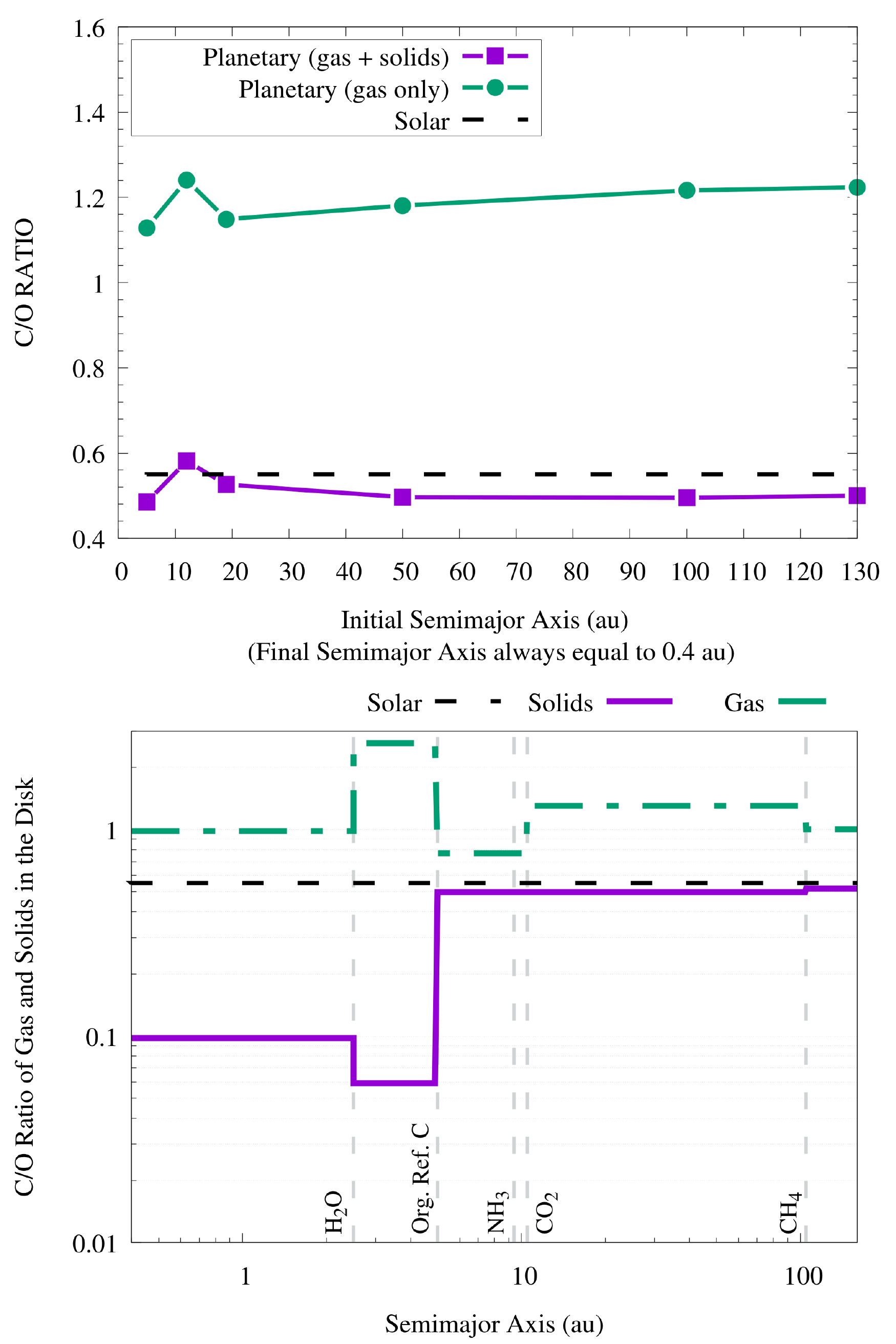}
    \caption{\textit{Top:} Final C/O ratios of the gaseous envelope of the giant planet as a function of its initial orbital position for the cases of solid-enriched giant planets (purple line and squares) and gas-dominated giant planets (green line and circles). \textit{Bottom:} C/O ratios of the gas (green dot-dashed line) and the solids (purple solid line) throughout the protoplanetary disk. The dashed lines shows the stellar C/O ratio, which is the same as the solar one. We refer the readers to Sect. \ref{sect-C_O_ratio} for details.}
    \label{fig-CO_ratio}
\end{figure}

The next step in our analysis has been to compute the overall C/O ratios of the giant planets at the end of the simulations. For comparison, we also computed the C/O ratios the same giant planets would have solely due to the gas they accreted, ignoring the contribution of the captured planetesimals. This second set of values can be physically interpreted as the C/O ratio of late-formed giant planets that migrated through protoplanetary disks already depleted of planetesimals by the passage of early-formed giant planets or by their radial drift induced by gas drag \citep[e.g.][]{turrini2015}. 

In the following, we will refer to solid-enriched giant planets when considering the contributions of both gas and solids, and to gas-dominated giant planets when considering only the contribution of gas. In the latter case, we will focus on elemental ratios involving only C, O, and N, as S is supplied only by planetesimals. Both sets of C/O ratios are reported in Table \ref{tab-elemental_ratios} and shown in the top panel of Fig. \ref{fig-CO_ratio}, where it is immediate to see that different and somewhat opposite behaviours characterize them.

The C/O ratio of gas-dominated giant planets is markedly super-stellar and has always values higher than one (see the green line and circle symbols in the top panel of Fig. \ref{fig-CO_ratio}). Conversely, the values of the C/O ratio of solid-enriched giant planets %is indistinguishable from the stellar one in most scenarios 
are close to the solar value and predominantly sub-solar (see the purple line and square symbols in the top panel of Fig. \ref{fig-CO_ratio}). These behaviours are a direct consequence of the distribution of C and O throughout the protoplanetary disk, % in our compositional model, 
particularly of the fact that the bulk of O and C gets trapped into solids within the first 2-5 au from the host star through rocks, H$_2$O ice, and refractory organic C (see Figs. \ref{fig-condensed_material} and \ref{fig-element_partitioning} and Sect. \ref{sect-compositional_model}). %\citealt{lodders2010,palme2014,bergin2015,mordasini2016,doyle2019,cridland2019})}.

The effects of this early trapping of the bulk of C and O in solids are illustrated by the bottom panel of Fig. \ref{fig-CO_ratio}, which shows the C/O ratios of the gas (green dot-dashed line) and solids (purple solid line) in the protoplanetary disk. Like the snowline of H$_2$O, the snowlines of refractory organic C and CO$_2$ are an order of magnitude closer to the star than that of CH$_4$: as a consequence, across most of the disk extension the C/O of the solids is characterized by slightly sub-solar values. In contrast, the C/O ratio of the gas is characterized by values around one, i.e. roughly twice the solar value.

As a result of the balance between accreted gas and solids (see Table \ref{tab-simulation_results}), the C/O ratio of solid-enriched giant planets  starting their formation beyond the CO$_{2}$ snowline (scenarios 3-6) monotonically decreases for increasing initial distances from the host star (or, equivalently, for increasing migration). Still, this variation is extremely limited (about 10\%, see Fig. \ref{fig-CO_ratio}). A similar yet opposite monotonic behaviour occurs for gas-dominated giant planets: their C/O ratios increase (also by about 10\% , see Fig. \ref{fig-CO_ratio}) for increasing initial distances from the star.

Both monotonic behaviours described above break once the giant planets start their formation close or inside the CO$_{2}$ snowline (scenarios 1 and 2). Both solid-enriched and gas-dominated giant planets show a peak in their C/O ratio in scenario 2 (giant planet starting to form at 12 au) due to a large amount of C-rich gas accreted between the snowlines of refractory organic C and H$_{2}$O (see Fig. \ref{fig-CO_ratio}, bottom panel). The C/O ratio then decreases again for giant planets starting to form within the NH$_{3}$ snowline. 

%The fact that both gas-dominated and solid-enriched giant planets show the same peak in the scenario where planetary formation starts at 12 au is a result of the sub-stellar amount of heavy elements captured through planetesimals by solid-enriched giant planets, which enhances the role played by the C contribution of the accreted gas.

%These behaviours are a direct consequence of the distribution of C and O throughout the disk in our solar nebula-like compositional model. The bulk of O and C gets trapped into solids within the first 2-5 au of the protoplanetary disk through rocks, H$_2$O ice, and refractory organic C (see Figs. \ref{fig-condensed_material} and \ref{fig-element_partitioning} and \citealt{lodders2010,palme2014,bergin2015,mordasini2016,doyle2019,cridland2019}), quickly bringing the C/O ratio of planetesimals close to the stellar value.

%\textbf{Outside} the CO$_{2}$ snowline, the main gaseous carriers of C and O in our template disk are CO and CH$_{4}$. As a result, the C/O ratio of the gas is slightly super-stellar, while that of solids is slightly sub-stellar. The C/O ratio of giant planets accreting most of their mass beyond the CO$_2$ snowline will reflect this duality, as shown by the results of scenarios 4-6 in Fig. \ref{fig-CO_ratio}. The values greater than one for the C/O ratio of gas-dominated giant planets in the regions within the CO$_2$ snow line, instead, are due to the trapping of about half the total O in rocks even \textbf{inside} the H$_2$O snowline \citep{lodders2010,palme2014, doyle2019}, which leaves similar amounts of C and O in the gas. 

Overall, our results suggest a dichotomy in the C/O ratio when giant planets form at large distances from their host stars, with values above one for gas dominated giant planets and about stellar for the solid-enriched ones. The fact that, for both classes of planets, the C/O ratio never varies by more than $10\%$ between different migration scenarios also suggests that, depending on the characteristics of the native protoplanetary disk, the diagnostic power of the C/O ratio might be limited. Specifically, it might not allow distinguishing between giant planets that formed closer or farther away from their stars, but only between gas-dominated or solid-enriched giant planets.

\subsection{Planetary migration and the C/N and N/O ratios}\label{sect-CNO_ratios}

\begin{figure*}
    \centering
    \includegraphics[width=\textwidth]{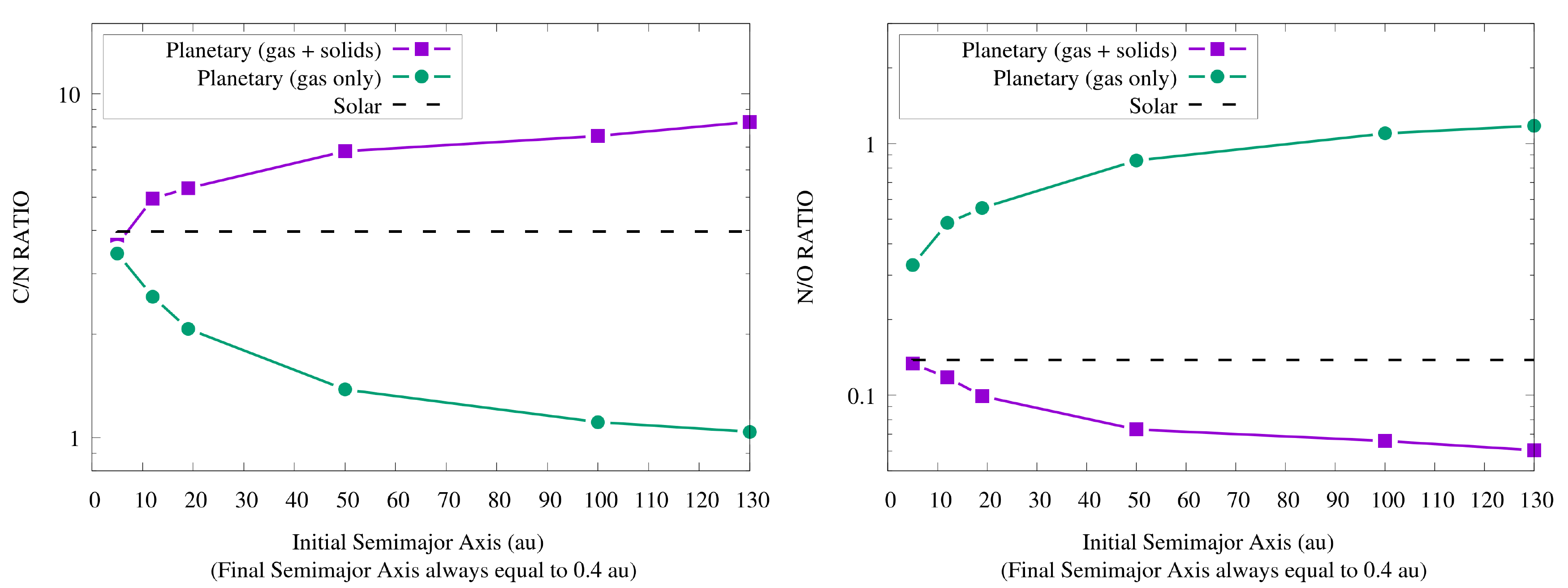}
    \caption{C/N (\emph{left}) and N/O (\emph{right}) ratios of the gaseous envelope of the giant planet as a function of its initial orbital position. In the figure we show the cases of solid-enriched giant planets (purple line and squares) and gas-dominated giant planets (green line and circles). The dashed lines show the stellar C/N and N/O ratios, which are the same as the solar ones. We refer the readers to Sect. \ref{sect-CNO_ratios} for details.}
    \label{fig-CNO_ratios}
\end{figure*}

\begin{figure*}
    \centering
    \includegraphics[width=\textwidth]{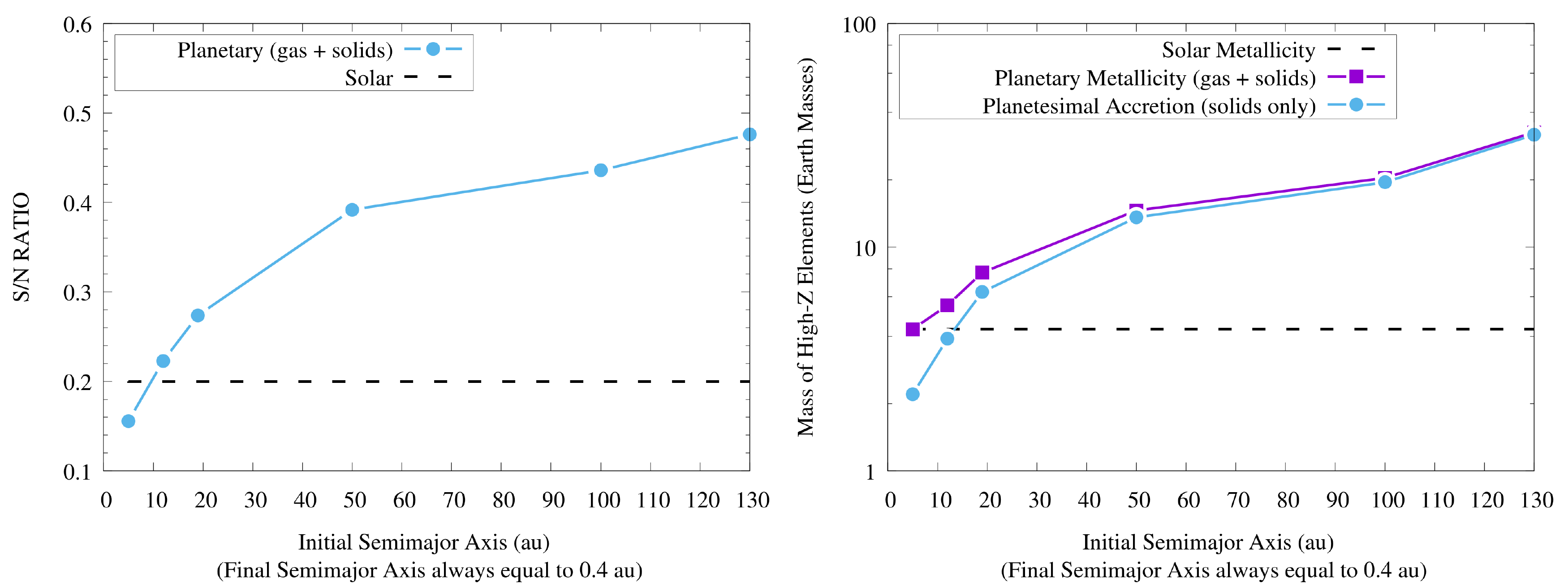}
    \caption{S/N ratio (\emph{left}), metallicity and mass of accreted planetesimals (\emph{right}) of solid-enriched giant planets as a function of their initial orbital positions. The dashed lines show the stellar S/N ratio and the mass of high-Z elements equivalent to the stellar metallicity, which are the same as the solar ones. We refer the readers to Sect. \ref{sect-SN_ratio} for details.}
    \label{fig-SN_ratio_and_enrichment}
\end{figure*}

Due to the possible limitations of the C/O ratio in discriminating between different formation regions and migration histories, we focused our attention on two ratios involving N, the third most abundant element among those we considered. Specifically, we considered the C/N and the N/O ratios. Their behaviour is shown in Table \ref{tab-elemental_ratios} and Fig. \ref{fig-CNO_ratios}, where we again consider the cases of solid-enriched and gas-dominated giant planets as discussed in Sect. \ref{sect-C_O_ratio}.

The C/N ratio of solid-enriched giant planets grows monotonically from a value of 3.6 in scenario 1 to a value of 8.3 in scenario 6 (see Table \ref{tab-elemental_ratios}, i.e. it grows with migration by about a factor of two. The C/N ratio of gas-dominated giant planets is characterized by the opposite trend. It grows smaller with migration by a factor of three, monotonically decreasing from a value of 3.4 in scenario 1 down to 1 in scenario 6.

The N/O ratio shows a similar but inverse behavior: the N/O ratio of solid-enriched giant planets decreases monotonically with migration. In contrast, the N/O ratio of gas-dominated giant planets increases monotonically. In the case of gas-dominated giant planets, the N/O ratio goes from a value of 0.3 in scenario 1 to a value of 1.2 in scenario 6, i.e. it grows by about a factor of four. For solid-enriched giant planets, the N/O ratio varies instead from a slightly sub-solar value of 0.13 in scenario 1 to about 0.06 in scenario 6, i.e. it decreases by a factor of two. 

It should be pointed out that the range of values discussed above is model-dependent, even if the overall behavior of the two ratios is not. The specific values of the C/N and N/O ratios depend on the partitioning of C, O and N among the different snowlines (see Fig. \ref{fig-element_partitioning}). As a result, the slope of the curves and the value of the C/N and N/O ratios will vary with the chemical and thermal profiles of protoplanetary disks. We will further explore this dependence in Sect. \ref{sec-alternative_N}.

%As an example, \citet{bosman2019} argue that the fraction of N in the form of NH$_3$ (33\%) adopted by \citet{eistrup2016,eistrup2018} and by our compositional model should be considered as an upper limit, and that more realistic partitioning would see about 90\% of N in the form of N$_2$. This would make the slopes in Fig. \ref{fig-CNO_ratios} steeper by a factor of a few, increasing the contrast in the C/N and N/O values between different scenarios.

Nevertheless, the trends discussed above depend exclusively on the fact that the bulk of N in disks is in the form of N$_2$ and that the N$_2$ snowline is farther away from the host star than all the C and O snowlines (see Fig. \ref{fig-element_partitioning}). While these two conditions are satisfied, the amount of C and O sequestered in solids will increase with the orbital distance faster than the amount of N, resulting in C/N and N/O trends similar to those shown in Fig. \ref{fig-CNO_ratios}.

The larger the deviation of the C/N and N/O ratios from the stellar values, therefore, the larger the orbital distance crossed by the giant planet during its disk-driven migration. Moreover, while the C/N ratio of both gas-dominated and solid-enriched giant planets might be observationally indistinguishable from the stellar one in the case of giant planets that experienced limited migration (see scenario 1 in the left-hand panel of Fig. \ref{fig-CNO_ratios}), the same does not apply to the N/O ratio.

A fraction of O varying between a quarter and half of its total abundance is already incorporated into rocks in the innermost regions of protoplanetary disks (\citealt{lodders2010,palme2014, doyle2019}, see Fig. \ref{fig-element_partitioning} and Sect. \ref{sect-compositional_model} for a discussion). Consequently, the N/O ratio of gas dominated giant planets will always be characterized by super-solar values (see scenario 1 in the right-hand panel of Fig. \ref{fig-CNO_ratios}). 

More generally, the comparison between Figs. \ref{fig-CO_ratio} and \ref{fig-CNO_ratios} highlights how the joint use of the three elemental ratios C/O, C/N\, and N/O allows to more easily break possible degeneracies in the interpretation of their values and their link with the formation and migration history of giant planets.

\subsection{Planetesimal enrichment and the S/N ratio}\label{sect-SN_ratio}

The final focus of our analysis has been the investigation of the S/N ratio as a direct tracer of the enrichment in heavy elements of giant planets. As highlighted by our results (see Sect. \ref{sect-enrichment} and Tables \ref{tab-simulation_results} and \ref{tab-simulation_gas_only}), both the metallicity and the composition of the gaseous envelope of giant planets are strongly affected by the enrichment in high-Z elements due to planetesimal accretion. %which therefore provides important indications on the extent of disk-driven migration undergone by giant planets. 

The planetary metallicity, however, coincides with the enrichment due to planetesimal accretion only in case of extensive migration of the giant planet. In contrast, in case of moderate or limited migration the two quantities increasingly diverge (see Table \ref{tab-simulation_results} and the comparison between the purple and blue lines in the right-hand panel of Fig. \ref{fig-SN_ratio_and_enrichment}). The planetary metallicity, moreover, is constrained indirectly through the planetary mass-radius relationship and, as such, its estimation is affected by the uncertainties not only on the measurements of both quantities but also on the interior modelling of giant planets. 

The investigation of the formation and migration histories of giant planets through their envelope metallicity would therefore benefit from additional and independent constraints on the enrichment in heavy elements due to planetesimal accretion. As shown by Fig. \ref{fig-SN_ratio_and_enrichment} and Tables \ref{tab-simulation_results} and \ref{tab-elemental_ratios}, the S/N ratio provides such an independent constraint, as its value monotonically grows with migration (Fig. \ref{fig-SN_ratio_and_enrichment}, left-hand panel) and is proportional to the mass of accreted planetesimals (see the blue curves in the left-hand and right-hand panels of Fig. \ref{fig-SN_ratio_and_enrichment}).

The proportionality between S/N and enrichment is not linear, as illustrated by the different scales of the S/N and planetesimal accretion curves as a function of migration (linear for the S/N values in the left-hand panel, logarithmic for the planetesimal accretion in the right-hand panel of Fig. \ref{fig-SN_ratio_and_enrichment}). This difference is a reflection of S representing a decreasing fraction per unit mass of the planetesimals for increasing distances from the host star. 

Specifically, inside the H$_2$O snowline planetesimals are composed of chondritic rock (see Fig. \ref{fig-condensed_material}), which is the main carrier of S (see Fig. \ref{fig-element_partitioning} and Sect. \ref{sect-compositional_model}). Moving away from the star, the composition of planetesimals will include increasing contributions of volatile materials (H$_2$O, CO$_2$, NH$_3$, etc., see Fig. \ref{fig-condensed_material}) not traced by S. As can be seen from Fig. \ref{fig-condensed_material} in the outer regions of the disk S will trace only half the mass (i.e. the rocky fraction) accreted through planetesimals. Consequently, while the proportionality coefficient between S/N and planetesimal accretion is constant within any given pair of snowlines (i.e. for a constant planetesimal composition), it changes each time a snowline is crossed (see Figs. \ref{fig-condensed_material} and \ref{fig-SN_ratio_and_enrichment}).
   
Nevertheless, the behavior of the S/N ratio as a tracer of the enrichment from planetesimal accretion shown in Fig. \ref{fig-SN_ratio_and_enrichment} depends only on the fact that, while the bulk of N in disks is in gaseous form as N$_2$ (see Fig. \ref{fig-element_partitioning} and Sect. \ref{sect-compositional_model} and \citealt{eistrup2016,eistrup2018,oberg2019,bosman2019}), the bulk of S is instead trapped in refractory solids (see Fig. \ref{fig-element_partitioning}, Sect. \ref{sect-compositional_model} and \citealt{lodders2010,palme2014,kama2019}). Consequently, even if the slope of the S/N curve as a function of planetary migration will depend on the NH$_3$-N$_2$ and the S refractory-volatile partitionings, the direct proportionality between S/N ratio, planetary migration and enrichment in high-Z elements due to planetesimal accretion will not. 

A close match between the planetesimal enrichment estimated through the S/N ratio and independent evaluations of the planetary metallicity, therefore, provides a strong indication that the giant planet under investigation formed at large distances from the host star and underwent extensive disk-driven migration. Conversely, a sub-stellar S/N ratio coupled with an estimated stellar or super-stellar planetary metallicity provides a clear indication that the giant planet experienced a moderate disk-driven migration and its metallicity is due to the joint contributions of gas and planetesimals.

\begin{figure*}
    \centering
    \includegraphics[width=\textwidth]{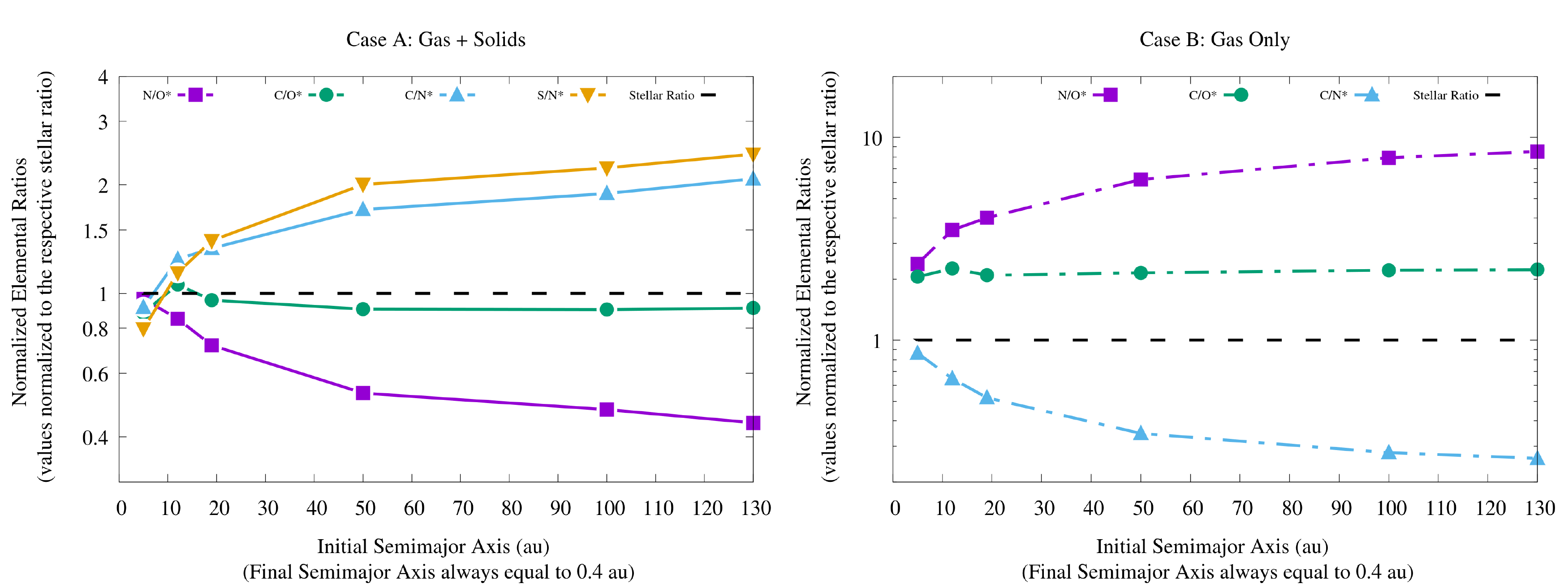}
    \caption{\emph{Left:} normalized elemental ratios of the gaseous envelope when the metallicity of the giant planet is dominated by the accretion of planetesimals. \emph{Right:} normalized elemental ratios in the gaseous envelope when the metallicity of the giant planet is dominated instead by the accretion of gas. Each elemental ratio is normalized to the relevant stellar elemental ratio.}
    \label{fig-all_ratio}
\end{figure*}

\section{Discussion}\label{sect-discussion}

\begin{table*}%[tb]
    \centering
    \caption{Normalized elemental ratios of the gaseous envelope}\label{tab-normalized_ratios}
    \begin{tabular}{c c c c c c c c c c}
    \hline
    Formation   & & \multicolumn{4}{c}{Gas + Solids} & & \multicolumn{3}{c}{Gas Only}\\
    \cline{3-6} \cline{8-10}
    Scenario    & &  C/O* & N/O* & C/N* & S/N* &  & C/O* & N/O* & C/N* \\
    \hline
    1	& &	0.88	&	0.97	&	0.91	&   0.80 & &	2.05	&	2.38	&	0.86  \\
    2	& &	1.06	&	0.85	&	1.24	&	1.14 & &	2.26	&	3.50	&	0.65  \\
    3	& &	0.96	&	0.72	&	1.33	& 	1.40 & &	2.09	&	4.02	&	0.52  \\
    4	& &	0.90	&	0.53	&	1.71	& 	2.00 & &	2.15	&	6.20	&	0.35  \\
    5	& &	0.90	&	0.48	&	1.89	& 	2.23 & &	2.21	&	7.94	&	0.28  \\
    6	& &	0.91	&	0.44	&	2.08	& 	2.44 & &	2.23	&	8.52	&	0.26  \\
    \hline
    \end{tabular}
    \justify
    \textbf{Note. }Elemental ratios of the gaseous envelopes of the giant planets normalized to the respective stellar values in the different formation scenarios when considering the contributions of both gas and solids and of gas only. A value of 1 indicates a stellar value of the elemental ratio.
\end{table*}

Our results show how the C/N, N/O and S/N ratios can be used as tracers of migration and planetary metallicity in studying the formation and dynamical history of individual giant planets, complementing the information provided by the C/O ratio. The different ranges of values they span, however, can hinder the efficient joint use of the four elemental ratios.

Moreover, the specific shapes of the curves of the four elemental ratios discussed in Sect. \ref{sect-results} depend on the characteristics of the host system and the thermo-chemical environment of its protoplanetary disk. Particularly critical, for the tracers we investigated, is the partitioning of N between NH$_3$ and N$_2$, as it determines the amount of N accreted with solids and, consequently, the slope of the C/N, N/O and S/N curves. 

In the following, we will show how the use of elemental ratios normalized to the respective stellar values allows for extracting information from the composition of giant planets and drawing general conclusions on their formation history even in absence of detailed information on their birth environment. To further test the validity of these normalized ratios, we will also investigate a compositional scenario characterized by an alternative NH$_3$:N$_2$ partitioning with respect to that of \citet{eistrup2016,eistrup2018} shown in Fig. \ref{fig-element_partitioning}.

%the comparison between giant planets belonging to different planetary systems, hence orbiting different host stars, is not necessarily straightforward when using the absolute values of these ratios. The extent of disk-driven migration and the associated enrichment in heavy elements of a giant planet can be constrained through the deviation of the elemental ratios from the respective stellar values, which will generally vary from star to star

\subsection{Normalized elemental ratios and their joint use}\label{sec-normalized_ratios}

To address the obstacle posed by the different ranges of values spanned by C/O, C/N, N/O and S/N, we investigated the use of normalized ratios, where each ratio is expressed in units of the corresponding stellar value. In this scale, giant planets whose composition matches the one of their host stars will possess C/N, N/O, and S/N ratios all equal to one. 

We recomputed the elemental ratios discussed in Sect. \ref{sect-results} in this normalized scale: the normalized ratios are reported in Table \ref{tab-normalized_ratios} while their resulting trends are shown in Fig. \ref{fig-all_ratio}, where we separate between gas-dominated (right-hand panel) and solid-enriched giant planets (left-hand panel). To avoid confusion between the normalized and non-normalized ratios, in the following we will refer to the normalized ratios as C/O*, N/O*, C/N*, and S/N*.

The direct benefit of this normalization is that the curves associated with the different elemental ratios can be more easily plotted together and compared. Their comparison immediately highlights additional information provided by the joint use of the C/O*, C/N*, and N/O* ratios. For gas-dominated giant planets, N/O* will be greater than C/O*, which will in turn be greater than C/N* (see Fig. \ref{fig-all_ratio}, left plot, and Table \ref{tab-normalized_ratios}).

For solid-enriched giant planets, the situation is inverted and C/N* will be greater than C/O*, which will, in turn, be greater than N/O* (see Fig. \ref{fig-all_ratio}, right plot and Table \ref{tab-normalized_ratios}). In both cases, the separation between the values of the three normalized ratios will increase with the extent of disk-driven migration experienced by the giant planet. In other words, the farther from the star a giant planet will start its formation, the more significant the difference between its C/N*, C/O*, and N/O* ratios.

The inclusion of the S/N* ratio in the comparison provides a further piece of information. Giant planets that experienced limited migration and enrichment in solid material will have C/N* ratios higher than the S/N* ratios (see Fig. \ref{fig-all_ratio}). This is a result of the comparatively large C contribution of the accreted gas, while S is supplied only by the limited amount of captured solids. Giant planets that experienced a significant migration and whose metallicity is dominated by the accretion of planetesimals will have S/N* ratios higher than the C/N* ratios (see Fig. \ref{fig-all_ratio}), as the gas cannot compensate for the lower accretion of C with respect to S with solids.

It is important to highlight how the relative behavior of the elemental ratios described above and shown in Fig. \ref{fig-all_ratio} is not an exclusive outcome of our compositional model. It is the general and direct result of the fact that, in protoplanetary disks, the bulk of S is incorporated in solids before the bulk of O and C, which in turn are incorporated into solids before the bulk of N. Furthermore, it is important to highlight that the role played by S in constraining the contribution of planetesimal accretion to the metallicity of giant planets is not unique to this element. As shown by \citet{turrini2018}, any element whose main carrier is characterized by a similar or lower volatility than S (or, more generally, O and C) can be used to the same effect.\\
\indent Consequently, the relative contributions of gas and planetesimals to the metallicity can be constrained by comparing the C/N* ratio with the normalized relative abundances with respect to N of rock-forming elements (e.g. Fe/N*, Si/N*, Mg/N*), refractory elements (e.g. Ti/N*, Ca/N*, Al/N*) or moderately volatile elements (e.g. Cl/N*, Cr/N*, Na/N*) as all these elements will be characterized by similar enrichment patterns (see \citealt{turrini2018} for a discussion, \citealt{lodders2010}, \citealt{palme2014}, \citealt{palme2017} for the observational and meteoritic perspective, as well as \citealt{kama2019} for the specific case of Ti).

\subsection{Implications of the NH$_3$:N$_2$ ratio in the disk}\label{sec-alternative_N}

\begin{figure*}
    \centering
    \includegraphics[width=\textwidth]{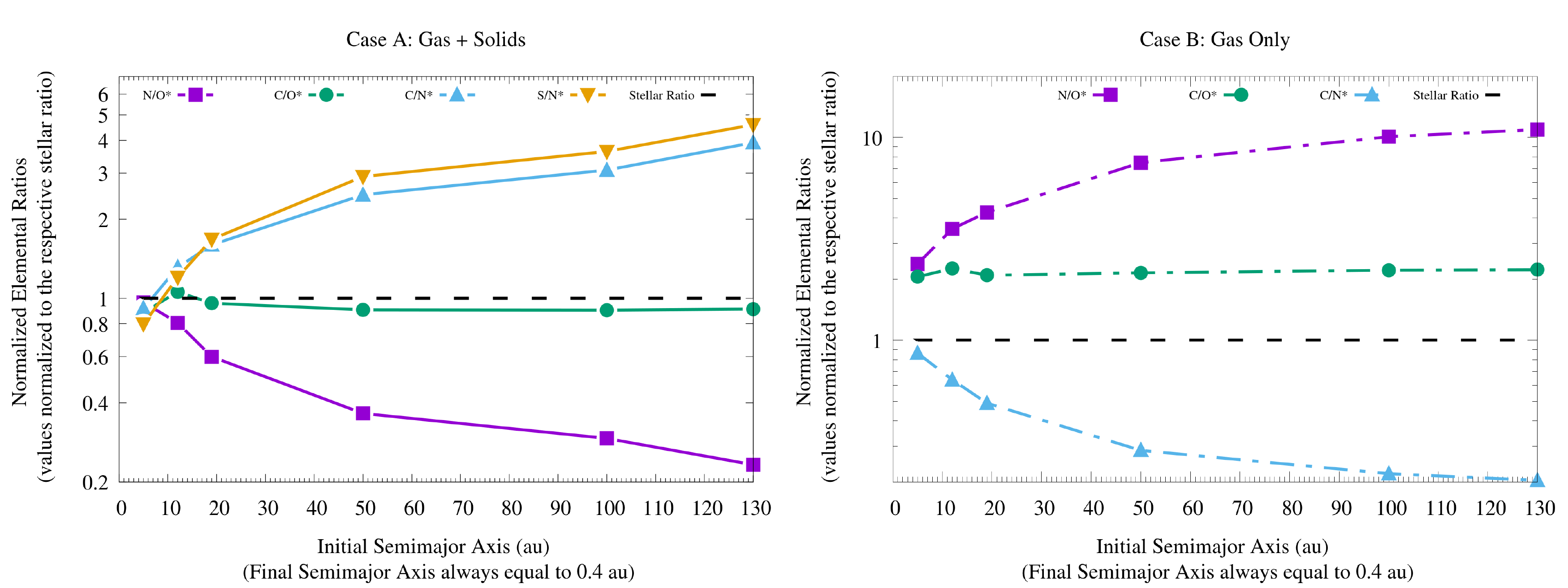}
    \caption{Same as Fig. \ref{fig-all_ratio} but assuming a 9:2 N$_2$:NH$_3$ ratio instead of the 1:1 ratio from \citet{eistrup2016,eistrup2018} adopted in our standard compositional model. The contrast among the different elemental ratios in the solid-enriched metallicity case is enhanced by a factor of two, while the effects of this alternative partitioning are more limited in the gas-dominated metallicity case.}
    \label{fig-alternative_N}
\end{figure*}

The partitioning of N between NH$_3$ and N$_2$ plays a central role in determining the N-based elemental ratios under investigation, both in their standard and normalized forms. The higher the fraction of N that remains in gaseous form as N$_2$, the steeper the slopes of the curves of the N-based elemental ratios. Recent work \citep{bosman2019} argued that the 1:1 NH$_3$:N$_2$ ratio adopted by astrochemical models of protoplanetary disks \citep{eistrup2016,eistrup2018} should be considered as an upper limit and, based on observational data, realistic partitioning should include significantly less ammonia.

To explore the impact of a lower abundance of NH$_3$ in protoplanetary disks, we modified our compositional model by changing the NH$_3$:N$_2$ ratio. We considered an alternative scenario where the NH$_3$:N$_2$ ratio is 2:9, which is equivalent to assuming that 90\% of N is in the form of N$_2$ and only 10\% is in the form of NH$_3$ (the major source of solid N in our compositional model). This alternative scenario falls into the range of values discussed by \citet{oberg2019} and \citet{bosman2019} and is consistent with the total N abundances measured by \textit{Rosetta} in the dust and ices of comet 67P C-G \citep[see][and references therein]{rubin2019}. The rest of the compositional model is kept unaltered. The normalized elemental ratios computed in this alternative compositional scenario are shown in Fig. \ref{fig-alternative_N}.

The only change between the alternative and standard compositional scenarios is in the slopes of the  curves of the N-based elemental ratios. The comparison of Figs. \ref{fig-all_ratio} and \ref{fig-alternative_N} reveals that the change is more marked in the case of solid-enriched giant planets, whose C/N* and S/N* ratios reach values twice as large than in the standard scenario for large-scale migration. Similarly, the N/O* ratio reaches values half as large than in the standard scenario. Analogous but more subdued changes are seen for gas-dominated giant planets, where the N/O* (C/N*) ratio increases (decreases) by $\sim$30\%. 

Overall, the comparison between the alternative and standard compositional scenarios shows that the global picture and the trends discussed in Sects. \ref{sect-results} and \ref{sec-normalized_ratios} do not depend on a specific partitioning of N among NH$_3$ and N$_2$ but solely of the fact that the bulk of N remains in gaseous form as N$_2$ for most of the extension of protoplanetary disks and condenses as ice farther away from the star than S, O, and C. 

The use of N-based elemental ratios as chemical tracers of the formation history of giant planets actually benefits from lower abundances of NH$_3$ in protoplanetary disks than those assumed in our standard compositional model. Lower NH$_3$ abundances enhance the contrast in the C/N* and N/O* ratios between different migration scenarios, allowing to put more accurate constraints on the formation regions and migration tracks of giant planets.

%The uncertainty in the partitioning of N between NH$_3$ and N$_2$ information provided by the comparison between the normalized C/O, C/N and S/N ratios we discussed in \ref{sec-normalized_ratios} is not influenced 

\subsection{Observational determination of the elemental ratios}
Our focus in this work has been on investigating the information content provided by the different elemental ratios on the formation history of giant planets. Such elemental ratios need to be retrieved from the atmospheric abundances of the main molecular carriers of C, O, N, and S by means of spectral observations \citep[see][and references therein for a discussion]{madhusudhan2016,madhusudhan2019}. Such observations will be systematically performed by the \textit{JWST} \citep{cowan2015,Greene2016} and \textit{Twinkle} \citep{edwards2018} and will be the main goal of the \textit{Ariel} mission \citep{tinetti2018}.\\
\indent The specific molecular carriers of C, O, N and S in a exoplanetary atmosphere will depend on its temperature and metallicity and on whether the atmospheric chemistry is governed by chemical equilibrium or disequilibrium \citep[e.g.][and references thereih]{madhusudhan2016,venot2018,madhusudhan2019, drummond2020}. For what it concerns S, the main molecular carrier is expected to be H$_2$S over a wide range of atmospheric temperatures \citep{fegley2010, woitke2018}. Under the assumption of chemical equilibrium, the main molecular carriers of C and O are expected to be H$_2$O and CH$_4$ for atmospheres colder than about 1200 K at the pressure of 1 bar and H$_2$O, CO, CO$_2$ for hotter ones \citep[e.g.][]{fegley2010,madhusudhan2016}.\\
\indent Similarly, the main carriers of N are expected to be NH$_3$ for atmospheres colder than about 700 K at the pressure of 1 bar and N$_2$ for hotter ones \citep{fegley2010,madhusudhan2016}. N$_2$ represents only a source of atmospheric opacity but cannot be identified directly by means of spectral observations, which would make the retrieval of the atmospheric abundance of N impossible for exoplanets hotter than 700 K. However, \citet{macdonald2017a,macdonald2017b} reported the detection of NH$_3$ and HCN in the atmospheres of hot Jupiters, suggesting an important role of disequilibrium chemistry \citep[see also][for a discussion]{venot2018} and the possibility to retrieve the abundance of N also at higher temperatures. Other studies also found potential evidence for those molecules in planets such as HAT-P-38\,b \citep{tsiaras2018,fisher2018} or HD-3167\,c \citep{guilluy2020}. However, those detections are very challenging with the \textit{Hubble Space Telescope}, as the molecular signals can be easily masked by water vapor and can be model dependent \citep{pinhas2018, changeat2020b}.    \\
\indent The atmospheric abundances of all these molecular species can be effectively retrieved with the JWST \citep{macdonald2017a,macdonald2017b,min2020} and Ariel \citep{tinetti2018,changeat2020,min2020} telescopes. In the specific case of Ariel, moreover, \citet{changeat2020} have recently shown that the abundances of H$_2$O, CO, CO$_2$, CH$_4$, and NH$_3$ can be retrieved even in case of markedly sub-solar abundances. While the observational implications of our results will be the subject of future dedicated investigations, retrieving the atmospheric abundances of H$_2$O, CO, CO$_2$, CH$_4$, NH$_3$, HCN, and H$_2$S should allow to directly estimate the elemental ratios discussed in this work and constrain the formation history of giant planets.

\subsection{Caveats and limitations on modelling and results}\label{sect-caveats}

%The more solid mass is incorporated into larger planetesimals, the more chemically decoupled the gas and solid phases of the disk are expected to become. 

Our modelling of the formation history of giant planets and of its compositional implications aims to capture the essential aspects of this complex process, and both the thermal and compositional structures of the protoplanetary disk in this work are based on data from the Solar System cross-calibrated with a number of extrasolar sources. As already reported in Sect. \ref{sect-model}, however, there are underlying assumptions and approximations that will require future investigations. We provide a brief discussion of their implications in the following.\\
\subsubsection{Characterization of the protoplanetary disk environment}
\indent Our model of the protoplanetary disk is a static one where both the thermal and the compositional structures are assumed as frozen over time. From a physical point of view this approximation can be viewed as equivalent to assuming that the conversion of dust and pebbles into planetesimals occurs rapidly on a timescale that is shorter than that over which the protoplanetary disk evolves chemically. %those of radial drift and chemical evolution of dust and gas. 
The sequestration of the bulk of the solid mass into planetesimals would cause the solid-gas interactions (e.g. gas-grain chemistry, condensation and desorption of ices) to become less effective due to the less favourable surface-to-volume ratio of planetesimals with respect to dust, slowing down the chemical evolution of the disk. This approximation is not unreasonable, based on the short timescales over which dust appears to be converted into planetesimals in disks \citep[e.g.][and references therein]{scott2007,manara2018} and the comparisons of the volatile budgets of comets and protostellar systems \citep[e.g.][]{bianchi2019,drozdovskaya2019}, but represents only one possible scenario. \\
\indent Protoplanetary disks are expected to evolve over time both in terms of their temperature profile and their distribution of the solid materials with respect to the gas. The cooling of the disk over time will result in an inward drift of the snowlines \citep[e.g.][]{panic2017,eistrup2018}, while the coupled interactions between radially drifting pebbles and dust grains and the thermal structure of the disk can result in the enrichment in heavy elements of the gas inward to the different snowlines due to the sublimation of the ices that cross the latter \citep{piso2016,booth2019,bosman2019}.\\
\indent While the implications of evolving disk environments on the compositional tracers here discussed will be addressed in future works, in the adopted compositional model the bulk of S, C and O are incorporated into low-volatility phases as refractories, organics and water ice. In parallel, the most volatile ices CO and N$_{2}$ never condense, while CH$_{4}$ accounts for a minimal part of the C and mass budgets (see Figs. \ref{fig-condensed_material} and \ref{fig-element_partitioning}). The only ice abundant enough in our compositional model to affect the chemical structure of the disk would be CO$_2$ (see Figs. \ref{fig-condensed_material} and \ref{fig-element_partitioning}). The inner location of its snowline, however, would cause the effects of CO$_2$ sublimation to be limited for giant planets accreting most of their mass at larger distances from the star.\\
\indent The sublimation-driven enrichment of heavy elements of the disk gas described by \citet{piso2016} and \citet{booth2019} is therefore going to be less pronounced in our template disk than in colder, more ice-rich disk scenarios \citep{bosman2019} and should not affect the results we described in a marked way. Moreover, as discussed in Sect. \ref{sec-normalized_ratios} the comparison between the C/N* and S/N* ratios would allow to constrain the relative contributions of gas and planetesimals to the planetary metallicity. The possibility of discriminating giant planets whose metallicity is dominated by planetesimal accretion from those where a significant fraction of the metallicity has been provided by the accretion of enriched nebular gas would allow to constrain the relative importance and frequency of the two processes.\\
\indent The inclusion of chemical reactions induced by the ionization environment to which the protoplanetary disk is exposed \citep{eistrup2016,booth2019}, particularly in compositional ``reset'' scenarios \citep{eistrup2016}, will have more marked effects on our results, due to the more complex distribution of the different molecular species and their snowlines. Additional differences could also arise from the roles played by molecular species not included in our model, such as O$_2$ as the main carrier of O in the inner and intermediate disk and by HCN as the main solid-phase carrier of N in the outer disk \citep{eistrup2016}. Also in this case, however, we would expect the large roles played by refractory and organic carriers of C, O and S and by gas as carrier of N to preserve the qualitative picture derived from our analysis.\\

\subsubsection{Characterization of planetary growth, migration and planetesimal capture}
\begin{figure}
    \centering
    \includegraphics[width=\hsize]{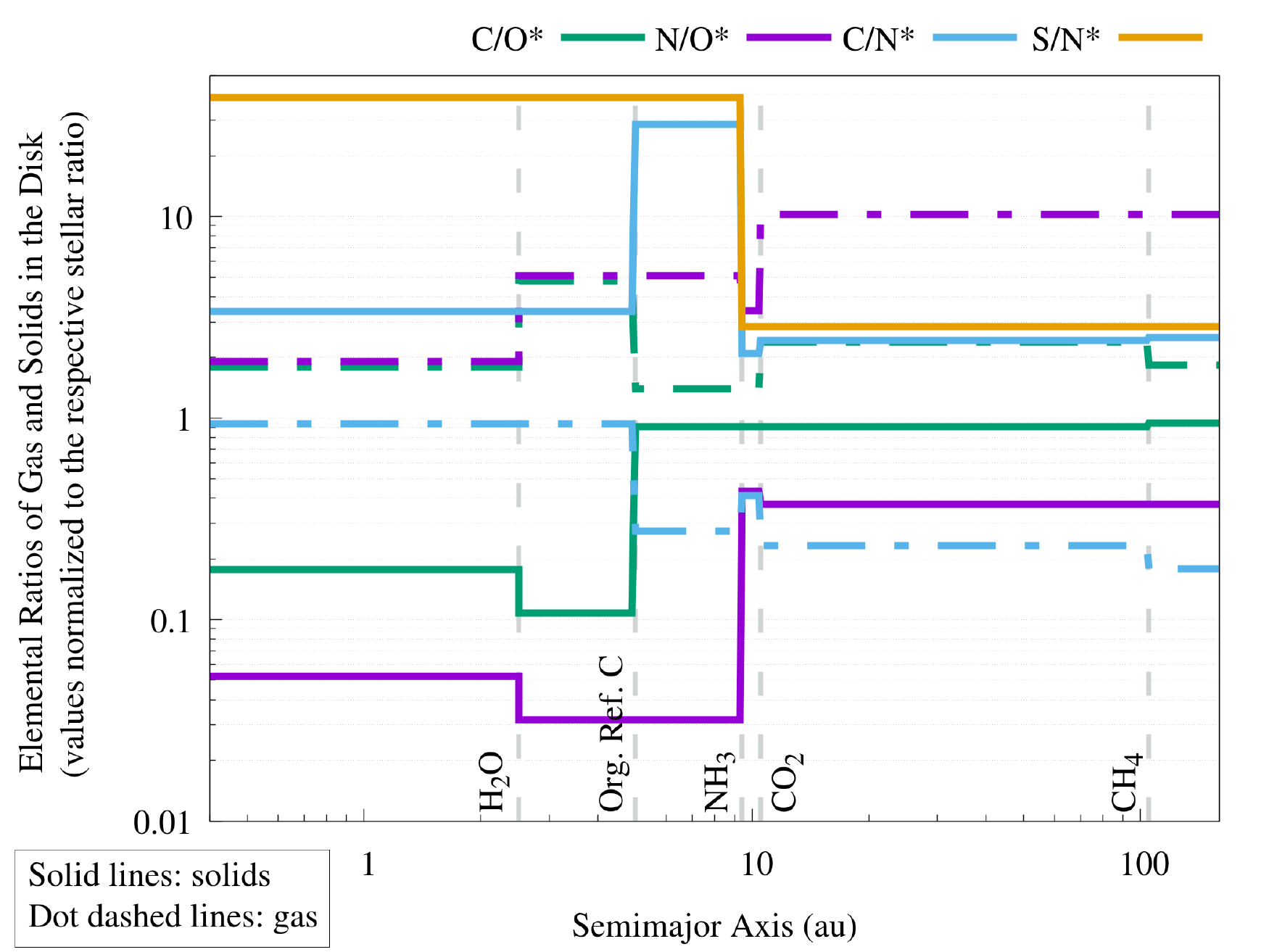}
    \caption{Elemental ratios of the gas (dot-dashed lines) and the solids (solid lines) in the protoplanetary disk as a function of the radial distance from the star. Each elemental ratio is normalized to the relevant stellar elemental ratio.}
    \label{fig-disk_elemental_ratios}
\end{figure}

The numerical approaches adopted for modelling the growth and migration of the giant planet we described in Sect. \ref{sect-giant_planets} provide an efficient way to explore the effects of different formation scenarios on the final planetary composition, as the change over time of the key quantities (mass and semimajor axis) depends only on their final values and the characteristic timescale of the involved process. This approach allows for capturing the fundamental aspects of the %growth and migration 
processes relevant to model the dynamical behaviour of the protoplanetary disk and the planetesimal accretion by the giant planet, but it is not physically self-consistent.\\
\indent The timescales of growth of the core and of the gaseous envelope and the migration rates during Type-I and Type-II migration are, in fact, dependent on each other and on the surrounding disk environment. Both the growth and the migration of the giant planet are linked to the characteristics of the disk environment and depend on parameters such as the disk viscosity, the gas accretion rate, the gas density profile, the pebble accretion rate and the disk scale height \citep[see e.g.][for recent discussions]{bitsch2015,bitsch2019,ida2018,ida2020,tanaka2020}.\\
\indent Furthermore, the formation scenarios we explored represent only a subset of the possible formation histories of giant planets and are focused specifically on the warm and hot giant planet populations that represent the ideal targets for spectroscopic investigations by future facilities like JWST and Ariel. Giant planets whose migration tracks end at larger distances from the host star, like in the case of the giant planets in the Solar System \citep[e.g.][]{bitsch2015,oberg2019,bosman2019} and of directly-imaged giant planets \citep{bowler2020}, will be characterized by different elemental ratios than their hot and warm counterparts even in case of similar bulk metallicities.\\
\indent Analogously, evolution tracks characterized by different values for the characteristic timescales $\tau_g$ in Eqs. \ref{eqn-collapsingradius} and \ref{eqn-typeIImigration}, where we adopted instead a common $\tau_g$ value, would imply a different evolution over time of the collisional cross-section of the migrating planet and, consequently, a different planetesimal accretion efficiency \citep[see the results of the parametric study by][for different migration rates and planetary masses/radii]{shibata2020}. Moreover, different migration timescales are expected to affect the orbital region where the giant planet accretes planetesimals most efficiently due to the balance between the effects of aerodynamic drag and resonant capture \citep{shibata2020}.\\
\indent Specifically, slower migration rates are expected to decrease the accretion efficiency in the inner regions of the disk crossed by the giant planets with respect to faster migration rates \citep{shibata2020}, as they provide more time for gas drag to act in regions where gas density is higher, with implications for the elemental ratios of the accreted material. Future investigations with more physically self-consistent treatments and larger sets of evolutionary tracks are therefore required to gain further insight on the compositional signatures of the formation process of giant planets.\\
\indent However, we can gain insight on the general validity of the results we described by inspecting the radial behaviour of the normalized elemental ratios of gas and solids in the protoplanetary disk as shown in Fig. \ref{fig-disk_elemental_ratios}. As first evidenced by the early work of \citet{oberg2011} in the case of the C/O ratio, the elemental ratios of gas and solids are characterized by opposite trends, with one possessing super-stellar values and the other sub-stellar values. While these trends are not strictly monotonic, the curves shown in Fig. \ref{fig-disk_elemental_ratios} never cross the value that represents the stellar elemental ratio.\\
\indent Consequently, while evolutionary tracks different from those we considered will be characterized by different absolute values of the planetary metallicity \citep[see also ][]{shibata2020} and elemental ratios with respect to the ones discussed in this work, the global picture depicted by Figs. \ref{fig-CO_ratio}-\ref{fig-disk_elemental_ratios} should be qualitatively preserved. \\
\indent Specifically, while the slopes of the curves in Figs. \ref{fig-CO_ratio}-\ref{fig-disk_elemental_ratios} may change, giant planets whose metallicity is dominated by planetesimal accretion will still possess normalized ratios C/N* $>$ C/O* $>$ N/O*, while giant planets whose metallicity is mainly contributed by the gas will possess N/O* $>$ C/O* $>$ C/N* (see Fig. \ref{fig-disk_elemental_ratios}). Furthermore, due to the link between metallicity and migration highlighted by the results of \citet{shibata2020} and this work, the differences between the values of these elemental ratios should still be proportional to the metallicity of the giant planet and, consequently, to the distance of its formation region from the host star.

\subsubsection{Distribution of the heavy elements inside the giant planet}
\begin{figure}
    \centering
    \includegraphics[width=\hsize]{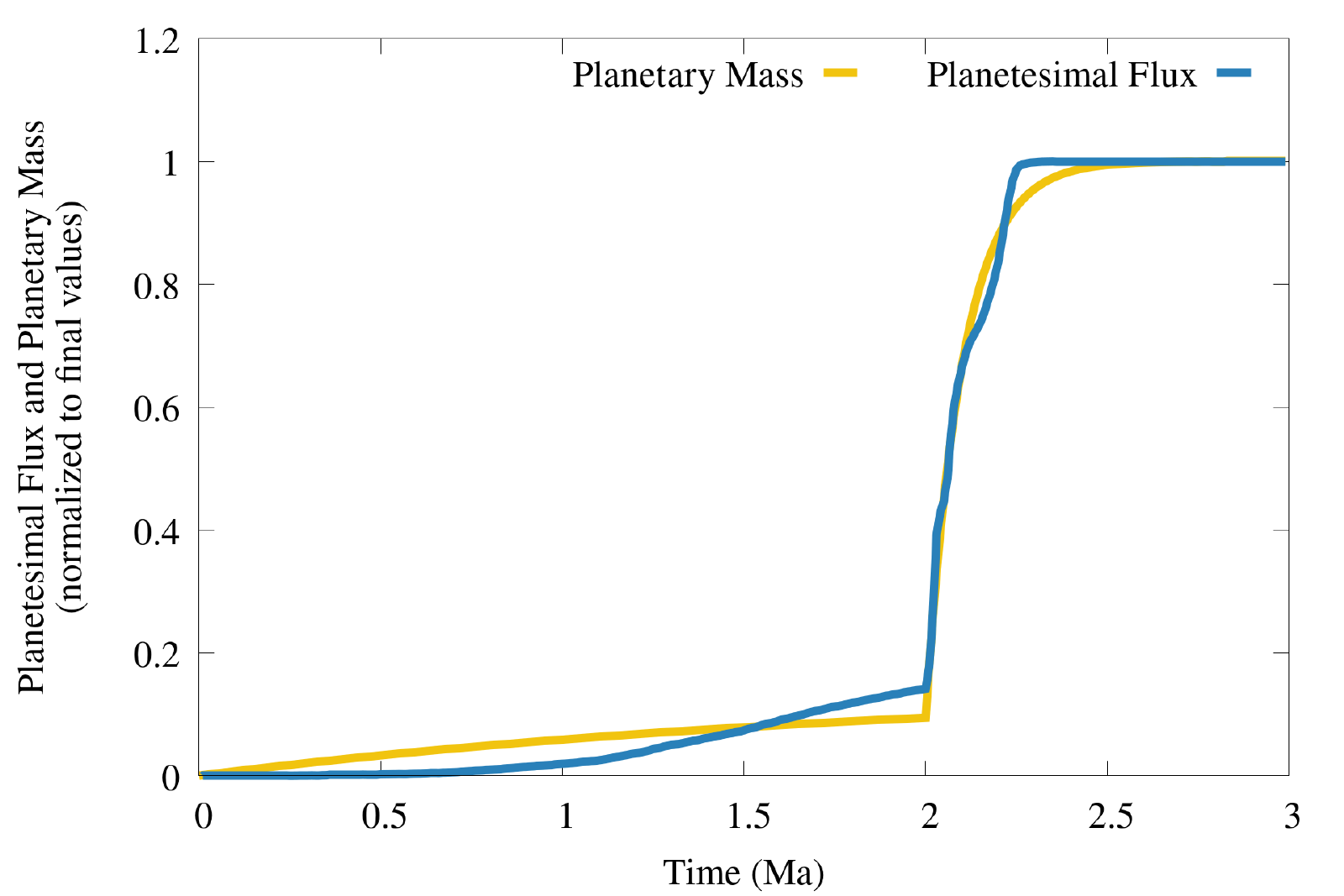}
    \caption{Mass growth and planetesimal accretion tracks, normalized to their final values, from the scenario where the giant planet starts forming at 19 au.}
    \label{fig-impact_flux}
\end{figure}

The elemental ratios we discussed have been computed under the assumption that all accreted heavy elements remain mixed within the gaseous envelope of the giant planet (see Sect. \ref{sect-planetesimal_disk} and \citealt{dangelo2014,mordasini2015,vazan2015,podolak2020}) and that the envelope and the atmosphere are compositionally homogeneous, either at the end of the formation process or after a few Gyr of interior evolution \citep[see Sect. \ref{sect-planetesimal_disk} and][]{vazan2015}. Furthermore, the core of the giant planet is implicitly assumed to remain compositionally distinct from the envelope and not to contribute to the envelope metallicity due to its erosion and dilution. Such scenario, however, is only one of the possible end states of the giant planet.\\
\indent It is possible that the interior of the giant planet is characterized instead by a layered structure, as recently suggested also for the case of Jupiter (see \citealt{debras2019} and \citealt{stevenson2020} and references therein for a discussion). If such layered structure is established early in the life of the giant planet, it may prevent the envelope and the atmosphere from becoming compositionally homogeneous and, as a result, cause them to be characterized by different elemental ratios. In other words, the atmosphere of the giant planet might reflect only the composition of the envelope outer layer, not of the giant planet as a whole. In this scenario, the atmospheric composition would constrain the gas and planetesimal accretion history of the giant planet during the tail of its formation process  (see \citealt{turrini2015} for a discussion) and provide a lower limit to the formation distance from the host star of the giant planet.\\
\indent Another possibility is that material from the core significantly contributes to the envelope metallicity of giant planets by erosion and dilution (see \citealt{coradini2010,helled2018,stevenson2020} and references therein, and \citealt{oberg2019} for an in-depth discussion of the compositional implications). In such a scenario and without significant planetesimal accretion occurring during migration, the atmospheric elemental ratios of the giant planet could trace the original formation region of the planetary core \citep{oberg2019}.\\
\indent However, the results of \citet{shibata2020} and this work indicate that a significant fraction of the envelope metallicity, if not the dominant one, arises from planetesimal accretion for giant planets experiencing large-scale migration while forming. As shown in Fig. \ref{fig-impact_flux} and discussed by \citet{podolak2020} for the case of non-migrating giant planets, the bulk of the planetesimal accretion occurs during the runaway gas accretion phase. As a result, the atmospheric elemental ratio would reflect the composition of the region where the giant planet accreted the bulk of its envelope and not that of the region, farther away from the host star, where the planetary core assembled (see also Fig. \ref{fig-planetary_tracks} for the comparison between the mass growth and migration tracks).

\section{Conclusions}\label{sect-conclusions}

Our work set out to investigate two main aspects of link between the formation process and the composition and elemental ratios of giant planets. The first one was the impact of formation regions extending far from the host star and migration tracks crossing multiple compositional regions of the protoplanetary disk. The second one was the new information 
provided by elemental ratios involving N and S, in addition to C and O. 

For a variety of initial orbital distances and migration tracks, we produced detailed n-body simulations of the interactions between growing and migrating giant planets and the surrounding planetesimal population. These simulations were coupled with a compositional model of the protoplanetary disk. The molecular abundances of the disk gas were inherited from the stellar core formation stage, while the elemental abundances in the planetary solids were calibrated on all available observational data (from meteorites, comets, and the Sun, to the interstellar medium, circumstellar disks, and polluted white dwarfs). 

Our results can be summarized as follows:
\begin{itemize}

\item We confirm the proportionality between planetary migration and metallicity discussed by \citet{shibata2020} in their study of the link between planetesimal accretion and migration. We show that the metallicity-migration correlation holds true also when the metallicity contribution of the accreted gas is accounted for, and shapes the abundance ratios of the high-Z elements of giant planets. This proportionality suggest that the higher the migration, the higher the planetesimal accretion efficiency of giant planets, and the more markedly super-stellar their metallicities. This, in turn, suggests that the spread in the metallicity data of Jovian planets discussed by \citet{thorngren2016}, \citet{wakeford2017} and \citet{sing2018} can be the reflection of different migration/planetesimal accretion histories.

\item The information provided by the C/O ratio is quite limited in case of giant planets acquiring the bulk of their C and O beyond the CO$_2$ snowline, where the main gaseous carriers of these elements are CH$_4$ and CO.\\ In the case of gas-dominated giant planets, the C/O ratio shows values greater than one, due to CH$_4$ increasing the C/O ratio of the gas with respect to the value of one otherwise set by CO. On the contrary, for solid-enriched giant planets the C/O ratio shows slightly sub-stellar values. \\ In both cases, however, the C/O ratio is essentially constant with the initial distance of the planetary seed, therefore providing little or no clue on the formation region. More marked changes in the C/O values are seen, in both cases, for giant planets growing within the CO$_2$ snowline, but even in these cases the C/O values do not vary by more than 10-20\%.\\ Such limited changes in the C/O values are smaller than the accuracy of current retrieval tools \citep[e.g.][]{barstow2020}: as a result, the C/O ratios of the different migration scenarios we considered in this work would be observationally indistinguishable from each other. Consequently, the C/O ratio would not allow to gain insight on the formation region and migration track of the giant planet.
 
\item The degeneracy in the information provided by C/O can be broken through the use of the C/N and N/O ratios, which provide stronger constraints on the formation region and extent of migration of giant planets (see Fig. \ref{fig-CNO_ratios}).\\ We find that the N/O ratio significantly grows with migration for gas-dominated/low metallicity giant planets and decreases for solid-enriched/high metallicity ones.\\ This is because the bulk of O is trapped into solids already in the inner regions of the protoplanetary disk, while the bulk of N remains in gas form until the outermost regions. The C/N ratio shows the opposite behaviour for similar reasons, i.e. due to the earlier incorporation of the bulk of C into solids with respect to N.

\item The behaviour of the C/N and N/O elemental ratios is fundamentally determined by the solidly established volatility of N, which causes the N content of giant planets to grow slower with planetesimal accretion than that of the other elements. The farther the giant planets start their migration from the host star, the more their C/N and N/O ratios  diverge from the stellar values.\\ For giant planets starting their formation increasingly close to the host star, the C/N ratio of giant planets will converge to the stellar value, as both C and N will be in gas form in the inner few au of protoplanetary disks (see Fig. \ref{fig-CNO_ratios}). 

\item The N/O ratios of gas-dominated and solid-enriched giant planets, however, will be significantly different even for very small migration distances, as a fraction of O is always trapped into rocks (see Fig. \ref{fig-CNO_ratios}).\\ Hence, if gas is the main source of high-Z elements for a giant planet, its N/O value will remain super-stellar. If the envelope of the giant planet is instead significantly enriched by solids, its N/O value will reach at most the stellar value.\\ Due to the monotonic trends of the planetary C/N and N/O ratios, giant planets accreting gas and solids in different regions of the disk will not share the same values of these elemental ratios even if they share the same overall metallicity.

\item The S/N ratio provides a direct window into the metallicity of giant planets. Specifically, given that the bulk of S is sequestered early into solids while the bulk of N remains in gas phase, the S/N ratio can be used as a proxy into the amount of solid material giant planets accrete as planetesimals during their formation history. This allows to validate the indirect estimates of the metallicity based on the mass-radius relationship of giant planets, but can in principle also allow to separate the metallicity contributions of gas and solids to obtain a more complete picture of the formation history of giant planets.

\item The C/O, C/N, N/O and S/N ratios normalised to their stellar values (indicated with the superscript *) reveal that solid-enriched giant planets will be characterized by C/N* $>$ C/O* $>$ N/O*. Gas-dominated giant planets, instead, will be characterized by N/O* $>$ C/O* $>$ C/N*. The farther away from the star the giant planet started its formation path, the greater the difference between the three normalized ratios. 

Moreover, giant planets for which planetesimal accretion is the main source of metallicity (as in the case of large-scale migration) will have S/N* $>$ C/N*. Giant planets for which both gas and solids contribute to the metallicity will have instead C/N* $>$ S/N*, with the contrast between the two values being higher the less the solids contribute to metallicity.\\ These trends depend on the fact that the bulk of S is incorporated in solids in protoplanetary disks before the bulk of O and C, which in turn are incorporated into solids before the bulk of N. As such, the comparison between S/N* and C/N* can be used to discriminate between high-metallicity giant planets that accreted nebular gas enriched in heavy elements \citep{booth2019} from those for which planetesimal accretion was the dominant source of heavy elements.

\item Finally, similarly to the case of the normalized metallicities values adopted by \cite{thorngren2016}, the use of normalized elemental ratios (not necessarily limited to C/N, N/O, C/O and S/N) removes the intrinsic compositional variability between different planetary systems due to the specific elemental budgets of their protoplanetary disks. This, in turn, opens up the possibility of more reliable comparisons between  the respective formation and migration histories of giant planets orbiting different stars.\\
Furthermore, any element whose main carrier is characterized by a lower or similar volatility than S can be used to compute normalized elemental ratios with respect to N and gain insight on the source of the planetary metallicity \citep{turrini2018}. The use of normalized elemental ratios therefore allows to compare the constraint on the metallicity derived for different giant planets using different low-volatility elements (e.g. S/N*,Al/N*,Na/N*,Cr/N*).

\end{itemize}

This first work illustrates the potential of elemental ratios involving sulphur and nitrogen in considerably expanding the toolkit available to probe the formation history of giant exoplanets. In subsequent works we will test the predictive power of our tracers against different protoplanetary disk chemical histories, and explore their application to exoplanetary atmospheres as probes into the formation history of exoplanets.

\acknowledgements
The authors wish to thank the two anonymous referees for their comments and suggestions that improved the presentation of the work and the discussion of its results and implications. The authors also wish to thank Nadia Balucani, Maria Teresa Capria, Cecilia Ceccarelli, Marco Fulle, Francesco Marzari, Yamila Miguel and Michiel Min, as well as the Planetary Formation Working Group of the Ariel Consortium, for useful discussions and comments during the development of this work.  D.T., E.S., S.F., S.M., R.P. and D.F gratefully acknowledge the support of the Italian Space Agency (ASI) through the ASI-INAF contract 2018-22-HH.0 and the Italian National Institute of Astrophysics (INAF) through the INAF Main Stream project ``ARIEL and the astrochemical link between circumstellar disks and planets'' (CUP C54I19000700005). D.F. acknowledges support from the Italian Ministry of Education, Universities and Research, project SIR (RBSI14ZRHR). M.K. was supported by the University of Tartu ASTRA project 2014-2020.4.01.16-0029 KOMEET, financed by the EU European Regional Development Fund. The computation resources for this work were supplied by the \textit{Genesis} cluster at INAF-IAPS. This research has made use of the Wolfram Alpha ``Computational Intelligence'' service and of the NASA Astrophysics Data System Bibliographic Services. This project has received funding from the European Research Council under the European Union's Horizon 2020 research and innovation programme (ERC grant agreement No 758892, ExoAI), the European Union's Seventh Framework Programme (FP7/2007-2013) and the ExoLight project (ERC grant agreement No 617119). Furthermore, the authors acknowledge funding by the Science and Technology Funding Council (STFC) grants: ST/K502406/1, ST/P000282/1, ST/P002153/1, ST/S002634/1 and ST/T001836/1. The project received support from the UCL London-Rome Cities Partnerships Program: ASI grant n. 2018.22.HH.O.

%-------------------------------------------------------------------

\bibliography{Bibliography.bib}
\bibliographystyle{aasjournal}

\end{document}